\documentclass[%
reprint,
superscriptaddress,
nofootinbib,
amsmath,amssymb,
aps,
pra,
]{revtex4-2}

\usepackage{enumerate}
\usepackage{subfigure}
\usepackage{mathrsfs}
\usepackage{graphicx}
\usepackage{dcolumn}
\usepackage{bm}
\usepackage[colorlinks=true, allcolors=blue]{hyperref}

\allowdisplaybreaks
\newcommand{\ket}[1]{| #1 \rangle}

\newcommand{\fullvec}[1]{\boldsymbol{\vec{#1}}}
\newcommand{\unitvec}[1]{\boldsymbol{\hat{#1}}}

\begin{document}


\title{Geometric mechanisms enabling spin- and enantio-sensitive observables in one photon ionization of chiral molecules}
\author{Philip Caesar M. Flores}
\affiliation{Max-Born-Institut, Max-Born-Str. 2A, 12489 Berlin, Germany}
\email{flores@mbi-berlin.de}
\author{Stefanos Carlstr\"om}
\affiliation{Max-Born-Institut, Max-Born-Str. 2A, 12489 Berlin, Germany}
\author{Serguei Patchkovskii}
\affiliation{Max-Born-Institut, Max-Born-Str. 2A, 12489 Berlin, Germany}
\author{Misha Ivanov}
\affiliation{Max-Born-Institut, Max-Born-Str. 2A, 12489 Berlin, Germany}
\affiliation{Insitute of Physics, Humboldt University zu Berlin, Berlin 12489, Germany}
\affiliation{Technion - Israel Institute of Technology, Haifa, Israel}
\author{Andres F. Ordonez }
\affiliation{Max-Born-Institut, Max-Born-Str. 2A, 12489 Berlin, Germany}
\affiliation{Department of Physics, Imperial College London, SW7 2BW London, United Kingdom}
\affiliation{Department of Physics, Freie Universit\"at Berlin, 14195 Berlin, Germany}
\author{Olga Smirnova}
\affiliation{Max-Born-Institut, Max-Born-Str. 2A, 12489 Berlin, Germany}
\affiliation{Technion - Israel Institute of Technology, Haifa, Israel}
\affiliation{Technische Universit\"at Berlin, 10623 Berlin, Germany}
\email{smirnova@mbi-berlin.de}
\date{\today}

\begin{abstract} 
We examine spin-resolved photoionization of randomly oriented chiral molecules via circularly polarized light, and revisit earlier predictions of Cherepkov (\href{https://iopscience.iop.org/article/10.1088/0022-3700/16/9/013/meta}{J. Phys. B: Atom. Mol. Phys. \textbf{16}, 1543, 1983}). We will show that the dynamical origin of spin- and enantio-sensitive observables arise from two intrinsic mechanisms that are quantified by two pseudovectors stemming from the geometric properties of the photoionization dipoles in spin space and in real space, and an extrinsic mechanism which is a directional bias introduced by the well-defined direction of light polarization. These mechanisms arise solely from electric dipole interactions. Consequently, this means that the ten independent parameters that was earlier predicted by Cherepkov to fully describe spin-resolved photoionization of chiral molecules can be reduced as moments of these three pseudovectors. We also find that the molecular pseudoscalars describing the spin- and enantio-sensitive components of the yield can be described by the flux of these pseudovectors through the energy shell, which changes sign upon switching enantiomers. Our results provide compact expressions for these observables which provide an intuitive picture on what determines the strength of these spin- and enantio-sensitive observables. The approach can be readily generalized to photoexcitation, multiphoton processes, and arbitrary field polarizations. Regardless of the specific driving conditions, the resulting spin- and enantio-sensitive observables are still controlled by the same three pseudovectors, underscoring their universal role as the primary generators of chirality-induced spin asymmetries, emphasizing their fundamental geometric origin and the universality of the mechanism identified here.
\end{abstract}

\maketitle


\section{Introduction}

Chirality plays a fundamental role in molecular physics, chemistry, and biology, manifesting itself as observables that change sign upon reversal of the molecular chirality. In photoionization, the most prominent example is photoelectron circular dichroism (PECD) wherein an ensemble of randomly oriented chiral molecules is ionized by circularly polarized light \cite{sparling_two_2025}. This results to a forward-backward asymmetry in the photoelectron angular distribution even within the electric-dipole approximation. PECD was first predicted for one-photon ionization by Ritchie in 1976 \cite{ritchie1976theory}, and later rediscoverd by Cherepkov in 1982 \cite{cherepkov1982circular}. Following early quantitative calculations of the expected effect \cite{powis2000}, the first measurement was done by B\"{o}wering   \cite{bowering} and was dramatically advanced in Refs. \cite{ulrich2008giant,nahon,nahon2015valence}. Subsequent demonstration of PECD in multiphoton ionization \cite{lux2012circular,lehmann2013imaging}, strong-field regimes \cite{beaulieu2016universality}, and other polarization geometries such as elliptically polarized light \cite{comby2018real} and  bichromatic laser fields \cite{Rozen2019PRX} cements the status of PECD as a truly universal probe \cite{sparling_two_2025}, and has been shown to yield strong enantiosensitive signals across several molecular species  \cite{garcia2003circular,turchini2004circular,hergenhahn2004photoelectron,garcia2008chiral,janssen,janssen2014detecting,powis2000photoelectron,stener2004density,artemyev2015photoelectron}.

While PECD reveals enantiosensitive asymmetries in the orbital angular distribution of the emitted electron, the photoelectron also carries an intrinsic spin degree of freedom. Even within the electric–dipole approximation, photoionization of chiral molecules can generate spin-polarized electrons without invoking magnetic interactions of the light field. Spin polarization by one photon ionization of  atoms was pioneered by Fano \cite{fano}, extended to resonant multiphoton ionization by Lambropolous \cite{lambropolous} and to tunneling ionization by Barth and Smirnova \cite{barth2013}. The interplay between molecular chirality, spin–orbit coupling, and the outgoing scattering dynamics thus opens an additional channel of enantiosensitive response. Moreover, the discovery of chirality induced spin selectivity (CISS) \cite{ray1999asymmetric}, wherein molecular chirality governs spin polarization, has reinvigorated the field and posed questions about the dynamical origins of spin-chirality coupling \cite{CISS}. Understanding how spin polarization emerges, and how it relates to the underlying electric-dipole transition amplitudes, provides deeper insight into the geometric and dynamical origin of chiral photoionization.

A general description for the spin-resolved one-photoionization of randomly oriented chiral molecules was pionered by Cherepkov \cite{cherepkov1983manifestations}, and showed that ten independent parameters are required to fully characterize all allowed spin and momentum resolved observables. Among these, five are only non-zero for chiral molecules. Meanwhile the rest are non-zero for both chiral and non-chiral molecules which have been studied theoretically and experimentally by several authors \cite{cherepkov1983spin,chandra1989photoelectron,chandra1989photoelectronTd,cherepkov1991comment,schonhense1984spin,heinzmann1981spin}. This establishes that spin-resolved photoionization of chiral molecules is a fundamentally richer problem than its achiral counterpart, while at the same raises the question on the dynamical origin of the resulting spin- and enantio-sensitive observables. 

In the case of PECD as well as other enantio-sensitive observables, these can be accessed via the multipolar expansion of the photoelectron angular distribution 
\begin{equation}
W( \unitvec{k} ) = \sum_{\ell,m_\ell} b_{\ell,m_\ell} Y_{\ell,m_\ell}(\unitvec{k}),
\end{equation}
where, $\unitvec{k}$ is the photoelectron momentum, and $Y_{\ell,m_\ell}$ are the complex-valued spherical harmonics. The non-zero coefficients $b_{\ell,m_\ell}$ will depend on the selection rules for a given setup and the enantio-sensitivity arises from the interference of the continuum partial waves embedded in them (see for example Ref. \cite{ordonez2022disentangling}). The dynamical origin of these enantiosensitive observables in one or multiphoton ionization is addressed by geometric magnetism \cite{ordonez2023geometric} (for the case of photoexcitation see Refs. \cite{ordonez2024temporal_geometry,roos2025temporal_geometry}). Its central object is the geometric propensity field: 
\begin{align}
	\fullvec{B}_{\fullvec{k}} \equiv i\fullvec{D}_{\fullvec{k}}^{*}\times\fullvec{D}_{\fullvec{k}},
	\label{eq:Bfield}
\end{align} 
where, $\fullvec{D}_{\fullvec{k}}$ is the photoelectron dipole field. The propensity field underlies several classes of such observables that originate from its multipole moments \cite{ordonez2023geometric}.  

Here, we revisit the pioneering work of Cherepkov \cite{cherepkov1983manifestations} and extend the methods developed in Ref. \cite{ordonez2022geometric} to include the photoelectron spin degree of freedom. We will show that the dynamical origins of the such observables arise from two intrinsic mechanisms that are quantified by two pseudovectors stemming from the geometric properties of the photoionization dipoles in spin space and in real space, and an extrinsic mechanism which is a directional bias introduced by the well-defined direction of light polarization. Specifically, we will show that the ten parameters earlier predicted by Cherepkov can be expressed as moments of these three pseudovectors, 

The main idea of the approach is to not invoke any partial wave expansion on the scattering wavefunction to obtain compact analytic expression that reveal how the geometry of the pseudovector fields maps onto the spin- and enantio-sensitive observables. Specifically, we find that these observables can be described as flux of effective pseudovector field through the surface of the energy shell. Moreover, we find that the spin- and enantio-sensitive observables can be as large or even larger than PECD thereby presenting new avenues for chiral discrimination. Note that our approach does not suggest that partial wave expansion is not needed to numerically calculate these parameters. Instead, we are mainly interested with the connection of these effective psuedovectors, that are quantified by the photoionization dipoles, to the spin- and enantio-sensitive photoionization parameters. The approach thereby reveals physics that are otherwise obscured by the partial wave expansion.

The rest of the paper is outlined as follows. In Sec. \ref{sec:rev-cherpkov}, we provide a brief review of Cherepkov's appraoch. In Sec. \ref{sec:vector-form} we closely follow Cherepkov's approach and adopt the vectorial formulation that was applied in Refs. \cite{ordonez2022geometric,ordonez2018generalized} to provide a unified description of spin-polarized chiral electric-dipole response. In Secs. \ref{sec:physical_picture} and \ref{sec:total_sp}, we use synthetic chiral argon \cite{flores2024_3} as toy model to provide a physical picture the geometric mechanisms and photoionization parameters. Last, we conclude in Sec. \ref{sec:conc}.

\section{Review of Cherepkov's method}
\label{sec:rev-cherpkov}

\begin{figure}[t !]
	\centering
	\includegraphics[width=0.35\textwidth]{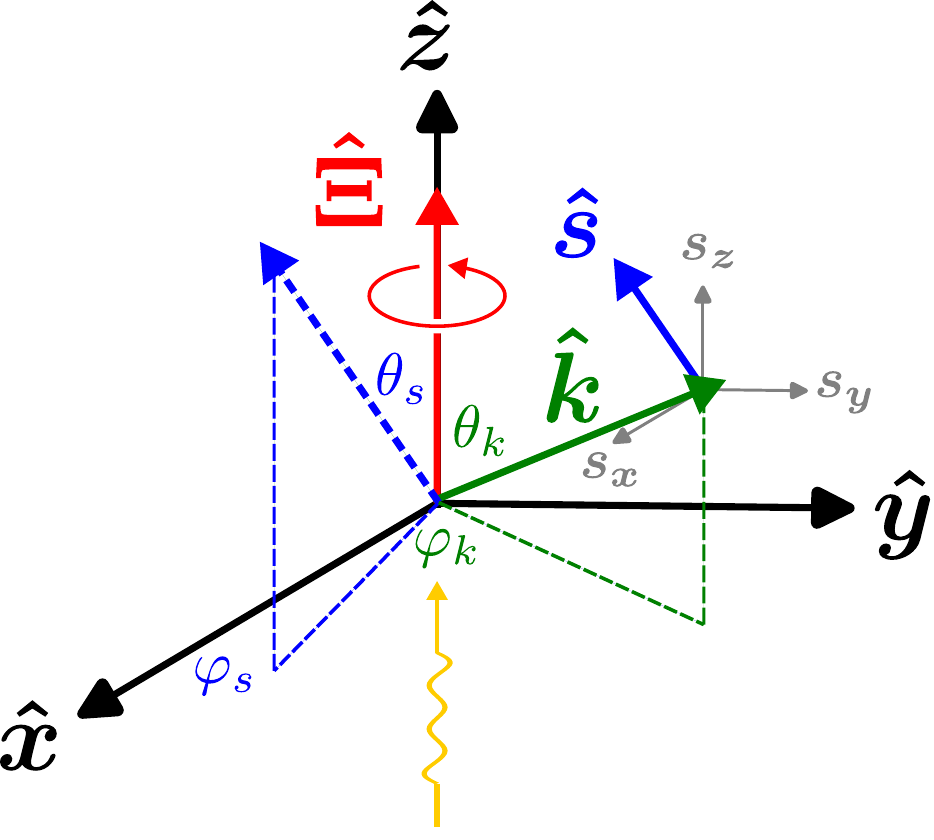}
	\caption{Specification of coordinates in the laboratory frame. The light field propagates along $\unitvec{z}$. The unit vector $\unitvec{\Xi}=\xi\unitvec{z}$ (solid red) is the direction of photon spin, where $\xi=\pm1$ is a dichroic parameter characterizing the direction of rotation of the light polarization vector. Upon ionization, the photoelectron is ejected in the direction of $\unitvec{k}$ (solid green) with its spin measured parallel to $\unitvec{s}$ (solid blue). The expansion $W^L(\unitvec{k}^L,\unitvec{s}^L)$, Eq. \eqref{eq:cherepkov-expansion} is represented by the the solid green and dashed blue vectors.} 
	\label{fig:setup}
\end{figure}

Using perturbation theory, the full spinor electron wave-function at the end of the ionizing pulse is\footnote{The superscripts $L$ and $M$ will be used to indicate quantities that are defined in the laboratory and molecular frame, respectively. Vectors that are defined in the molecular frame are transformed to the laboratory frame via the Euler rotation matrix $R_\rho$, i.e., $\fullvec{a}^L=R_\rho\fullvec{a}^M$.}: 
\begin{equation}		
    |\psi\rangle=|\psi_{o}\rangle+
		\sum_{I,\mu^L}
		\int\mathrm{d}\Theta_k^M\, c_{I,\fullvec{k}^M,\mu^L}
		|I\Psi_{I,\fullvec{k}^M,\mu^L}^{(-)}\rangle , 
		\label{eq:psi}
	\end{equation}
where, $I$ denotes the ion channel, $|\Psi_{I,\fullvec{k}^M,{\mu^L}}^{(-)}\rangle$ are fully spin-orbit coupled continuum states which are the two components of a spinor-valued scattering solution with opposite projections of spin (${\mu^L}=\pm\frac{1}{2}$) onto the laboratory z-axis $\unitvec{z}^L$, and
subject to the orthogonality condition
\begin{align}
\langle I_1\Psi_{I_1,\fullvec{k}_{1}^M,\mu_{1}^L}^{(-)}|I_2\Psi_{I_2,\fullvec{k}_{2}^M,\mu_{2}^L}^{(-)}\rangle=\delta_{I_1,I_2}\delta(\fullvec{k}_{1}^M-\fullvec{k}_{2}^M)\delta_{\mu_{1}^L,\mu_{2}^L}.
\end{align}
For one-photon ionization: 
\begin{subequations}
\begin{equation}
    c_{I,\fullvec{k}^M,\mu^L} =
    i\left( \fullvec{D}_{I,\fullvec{k}^M,\mu^L}^L\cdot\fullvec{E}^L \right),
\end{equation}
\begin{equation}
\fullvec{D}_{I,\fullvec{k}^M,\mu^L}^L=\langle I\Psi_{I,\fullvec{k}^M,\mu^L}^{(-)}|\fullvec{d}^L|\psi_0\rangle
\end{equation}
\end{subequations}
where, $\fullvec{D}_{I,\fullvec{k}^M,\mu^L}^L$ is the transition dipole matrix element, and $\fullvec{E}^L$ is the light field. Upon ionization, the photoelectron is ejected in the direction $\unitvec{k}^L$ while its spin is measured along the direction of $\unitvec{s}^L$ (see Fig. \ref{fig:setup}). 

Consider a circularly polarized light 
\begin{align}
\fullvec{E}^{L}=\frac{E_\omega^L}{\sqrt{2}}(\unitvec{x}^L+i\xi\unitvec{y}^L),
\end{align}
where, $\xi=\pm1$ is a dichroic parameter characterizing the direction of rotation of the light polarization vector.
To obtain the momentum- and spin-resolved photoionization yield, the dipole operator is first transformed from the laboratory to the molecular frame, i.e., 
\begin{align}
\fullvec{d}^L\cdot\fullvec{E}^L =& \sqrt{\dfrac{4\pi}{3}}r^L E_\omega^L Y_{1,\xi}(\Theta^L) \nonumber \\
=& \sqrt{\dfrac{4\pi}{3}}r^L E_\omega^L \sum_{\xi'} \mathscr{D}^1_{\xi',\xi} Y_{1,\xi'}(\Theta^M),  
\label{eq:dipole}
\end{align}
where, $\mathscr{D}_{m_j,m_j'}^{J}$ is the Wigner-D matrix. Next, the spatial part of the photoelectron wavefunction is transformed from the molecular frame to the laboratory frame while the spin function the other way around, i.e., 
\begin{align}
&\ket{I \Psi_{I,\fullvec{k}^M,\mu^L}^{(-)}} = \ket{I \psi_{I,\fullvec{k}^M,\mu^L}^{(-)}} | \chi_{\mu^L} \rangle  \nonumber \\
%
&= \sum_{\ell, m' } \left|I \varphi_{I,\ell,m',\mu^L}^{(-)} Y_{\ell,m'}^*(\unitvec{k}^M)\right\rangle | \chi_{\mu^L} \rangle  \nonumber \\
&= \sum \mathscr{D}^{1/2}_{\mu^M,\mu^L} \mathscr{D}_{m',m}^{\ell} Y_{\ell,m}^*(\unitvec{k}^L)  | I \varphi_{I,\ell,m',\mu^M}^{(-)} \rangle  | \chi_{\mu^M} \rangle.
\label{eq:photoelectron}
\end{align}
The first line follows from the assumption that the photoelectron wavefunction $|\Psi_{I,\fullvec{k}^M,\mu^L}^{(-)}\rangle$ can be factorized into its spatial component $|\psi_{I,\fullvec{k}^M,\mu^L}^{(-)}\rangle$ and spinor component $|\chi_{\mu^L}\rangle$ in the asymptotic scattering region. Meanwhile, the second line presents the standard partial wave expansion of the spatial component of the photoelectron wavefunction, and the last line imposes that the phase of $\varphi_{I,\ell,m',\mu^M}^{(-)}$ must depend on the spin-projection $|\chi_{\mu^M}\rangle$.
The primed and unprimed indices indicate values in the molecular and laboratory frame, respectively.

The photoionization yield for a given molecular orientation $W^L(\unitvec{k}^L,\unitvec{s}^L,\rho)$ is then obtained by taking the average of the spin-projection operator $\mathbb{\unitvec{P}}_{\unitvec{s}^{L}}$:
\begin{subequations}
\begin{align}
W^L(\unitvec{k}^L,\unitvec{s}^L,\rho)
=  \dfrac{4\pi}{3} |E_\omega^L|^2 & \sum \langle I \Psi_{I,\fullvec{k}^M,\mu_2^L}^{(-)}  | rY_{1,\xi} | \psi_o \rangle  \nonumber \\
 \mathbb{\unitvec{P}}_{\unitvec{s}^L} & \langle \psi_o | rY_{1,\xi}^* | I \Psi_{I, \fullvec{k}^M,\mu_1^L}^{(-)} \rangle ,
\label{eq:cherepkov_yield_rho} 
\end{align} 
\begin{equation}
\mathbb{\unitvec{P}}_{\unitvec{s}^{L}}=\frac{1}{2}(\mathbb{I}+\unitvec{s}^{L}\cdot\unitvec{\sigma}^L)
\end{equation}
\end{subequations}
where, $\mathbb{\unitvec{P}}_{\unitvec{s}^{L}}$ is the spin projection operator with respect to $\unitvec{s}^L$, and $\unitvec{\sigma}^L$ is the vector of Pauli spin matrices. Equation \eqref{eq:cherepkov_yield_rho} is then averaged over all molecular orientations  
\begin{align}
W^L  (\unitvec{k}^L,\unitvec{s}^L) = \int d\rho W^L(\unitvec{k}^L,\unitvec{s}^L,\rho)
\label{eq:cherepkov_yield}
\end{align}
in which, $\rho\equiv\alpha\beta\gamma$ are the Euler angles, and $\int d\rho \equiv \frac{1}{8\pi^2}\int_0^{2\pi}d\alpha \int_0^\pi d\beta \sin\beta \int_0^{2\pi}d\gamma$. 

\begin{widetext}
Using Eqs. \eqref{eq:dipole}-\eqref{eq:cherepkov_yield}, Cherepkov obtained the momentum and spin-resolved photoionization yield in a form reflecting the kinematic properties of the photoelectrons
\begin{subequations}\label{eq:cherepkov_Y}
\begin{align}
W^{L}(\unitvec{k}^{L},\unitvec{s}^{L})= & \sum A_{L,M_{L},S,M_{S}}Y_{L,M_{L}}(\unitvec{k}^L)Y_{S,M_{S}}(\unitvec{s}^L),
\label{eq:cherepkov-expansion} 
\end{align} 
\begin{align}
A_{L,M_{L},S,M_{S}}= & \dfrac{4\pi \sqrt{2\pi}}{3}  |E_\omega^L|^{2}  \sum(-1)^{m_{2}'+\xi'-\xi+\mu_{2}'-1/2} (2J+1) \sqrt{\frac{(2\ell_1+1)(2\ell_2+1)(2L+1)}{4\pi}} \nonumber \\
&\times 
\begin{pmatrix}
	\ell_{2} & \ell_{1} & L\\
	0 & 0 & 0
\end{pmatrix}
\begin{pmatrix}
	\ell_{2} & \ell_{1} & L\\
	m_{2}' & -m_{1}' & M_{L}'
\end{pmatrix} 
\begin{pmatrix}
	1 & 1 & J\\
	\xi'' & -\xi' & -M_{J}'
\end{pmatrix}
\begin{pmatrix}
	1 & 1 & J\\
	\xi & -\xi & 0
\end{pmatrix} \nonumber \\
&\times 
\begin{pmatrix}
	\frac{1}{2} & \frac{1}{2} & S\\
	\mu_{2}' & -\mu_{1}' & M_{S}'
\end{pmatrix}
\begin{pmatrix}
	L & S & J\\
	M_{L}' & M_{S}' & M_{J}'
\end{pmatrix}
\begin{pmatrix}
	L & S & J\\
	-M_{L} & -M_{S} & 0
\end{pmatrix}
(D_{\xi'}^{\ell_1,m_1',\mu_1'})^* D_{\xi''}^{\ell_2,m_2',\mu_2'},
\label{eq:cherepkovcoeff}
\end{align}
\end{subequations}
wherein, the quantities enclosed by large parenthesis are Wigner $3j$-symbols. Here, $D_{\xi'}^{\ell,m',\mu'} = \langle I \varphi_{I,\ell,m',\mu'}^{(-)} | rY_{\xi'} | \psi_i \rangle$ is the reduced transition dipole matrix element which is now completely defined in the molecular frame. 
\end{widetext}

The nonzero values for $A_{L,M_{L},S,M_{S}}$ are deduced from the properties of the $3j$-symbols, i.e., $0\leq L\leq3$, $0\leq S\leq 1$, $0\leq J \leq 2$, and $M_L+M_S=0$. Explicit evaluation of $A_{L,M_{L},S,M_{S}}$ will lead to thirteen terms in Eq. \eqref{eq:cherepkov-expansion}, however, only ten are relevant because of the relations:
\begin{subequations}
	\begin{equation}
		A_{2,1,1,-1}+A_{2,-1,1,1}= -\sqrt{3}A_{2,0,1,0}
	\end{equation}
	\begin{equation}
		A_{3,1,1,-1}=A_{3,-1,1,1}=-\sqrt{\frac{2}{3}}A_{3,0,1,0}.
	\end{equation}
\end{subequations}

The spherical harmonics $Y_{L,M_{L}}(\unitvec{k}^L)Y_{S,M_{S}}(\unitvec{s}^L)$ are then expressed in vector form such that Eq. \eqref{eq:cherepkov-expansion} becomes
\begin{widetext}
\begin{align}
	W^{L}(\unitvec{k}^{L},\unitvec{s}^{L})= \dfrac{\sigma_{\text{cross}}}{8\pi} & \left\{ 1 - \dfrac{\beta}{2} \left[3(\unitvec{k}^L\cdot\unitvec{\Xi}^L)^{2}-1\right] + A (\unitvec{s}^L\cdot\unitvec{\Xi}^L) - \eta (\unitvec{\Xi}^L\cdot\unitvec{s}^L\times\unitvec{k}^L)(\unitvec{k}^L\cdot\unitvec{\Xi}^L) \right. \nonumber \\
	&- \gamma \left[\frac{3}{2}(\unitvec{k}^L\cdot\unitvec{s}^L)(\unitvec{k}^L\cdot\unitvec{\Xi}^L)-\frac{1}{2}(\unitvec{s}^L\cdot\unitvec{\Xi}^L)\right] + D (\unitvec{k}^L \cdot \unitvec{\Xi}^L) + C (\unitvec{\Xi}^L\cdot\unitvec{s}^L\times\unitvec{k}^L)   \nonumber \\
	&+ \left. B_1 (\unitvec{k}^L \cdot \unitvec{s}^L) + B_2 (\unitvec{k}^L \cdot \unitvec{\Xi}^L) (\unitvec{s}^L \cdot \unitvec{\Xi}^L) + B_3 (\unitvec{k}^L\cdot\unitvec{\Xi}^L)^{2}(\unitvec{k}^L\cdot\unitvec{s}^L) \right\}
	\label{eq:kinematic}
\end{align}
\begin{subequations}\label{eq:parameters}
	\begin{minipage}{0.4\textwidth}
		\begin{align}
			\sigma_{\text{cross}} =& 2 \, A_{0,0,0,0} \label{eq:cherepkov-sigma}
		\end{align}
		\begin{align}
			\beta =& -\sqrt{5} \, \frac{A_{2,0,0,0} }{A_{0,0,0,0}} \label{eq:cherepkov-beta} 
		\end{align}
		\begin{align}
			A =& \sqrt{3} \, \frac{A_{0,0,1,0}}{A_{0,0,0,0}} \label{eq:cherepkov-A}
		\end{align}
		\begin{align}
			\eta=&i\frac{3\sqrt{5}}{2} \, \frac{(A_{2,-1,1,1}-A_{2,1,1,-1})}{A_{0,0,0,0}} \label{eq:cherepkov-eta}
		\end{align}
		\begin{align}
			\gamma =&-\sqrt{15} \, \frac{A_{2,0,1,0}}{A_{0,0,0,0}} \label{eq:cherepkov-gamma}  
		\end{align}
	\end{minipage}
	\begin{minipage}{0.6\textwidth}
		\begin{align}
			D=&\sqrt{3} \, \frac{A_{1,0,0,0}}{A_{0,0,0,0}} \label{eq:cherepkov-D}    
		\end{align}
		\begin{align}
			C =& -i \frac{3}{2} \, \frac{(A_{1,-1,1,1}-A_{1,1,1,-1})}{A_{0,0,0,0}} \label{eq:cherepkov-C}    
		\end{align}
		\begin{align}
			B_1 =& -\dfrac{3}{2} \left[\dfrac{(A_{1,-1,1,1}+A_{1,1,1,-1})+\sqrt{\frac{7}{3}}A_{3,0,1,0}}{A_{0,0,0,0}}\right] \label{eq:cherepkov-B1}
		\end{align}
		\begin{align}
			B_2 =& \dfrac{3A_{1,0,1,0} + \frac{3}{2}(A_{1,-1,1,1}+A_{1,1,1,-1}) - \sqrt{21}A_{3,0,1,0}}{A_{0,0,0,0}} \label{eq:cherepkov-B2}    
		\end{align}
		\begin{align}
			B_3 =& \frac{5\sqrt{21}}{2} \, \frac{A_{3,0,1,0}}{A_{0,0,0,0}} \label{eq:cherepkov-B3}
		\end{align}
	\end{minipage}
\end{subequations}
\end{widetext}
where, $\unitvec{\Xi}^L=-i(\fullvec{E}^{L*}\times\fullvec{E}^L)/|\fullvec{E}^L|^2=\xi\unitvec{z}^L$ is the direction of photon spin\footnote{There are errors for the parameters $\{\eta,D,C,B_1,B_2\}$ in the original expressions of Cherepkov (see Eq. 9 of Ref. \cite{cherepkov1983manifestations}) which we have corrected here. The relation of $\eta$ to $A_{L,M_{L},S,M_{S}}$ is not listed in Ref. \cite{cherepkov1983manifestations} but can be found in Eq. 87 of Ref. \cite{cherepkov1983spin}. Specifically, $\eta$ and $C$ had wrong signs, $D$ was missing the factor $\sqrt{3}$, $B_1$ should have the term $+\sqrt{7/3}A_{3,0,1,0}$, and $B_2$ should have $+3A_{1,0,1,0}$}. See Appendix \ref{app:cherepkov_yield} for details.

The parameters $\{\beta, A, \eta, \gamma \}$ are non-zero for both atoms and molecules, and the explicit form for the atomic case are provided in Ref. \cite{cherepkov1981theory}. Meanwhile, the parameters $\{D,C,B_1,B_2,B_3\}$ are only non-zero for chiral molecules. Moreover, the contribution of the parameters $\{ A, \gamma, D , C \}$ vanishes for linearly polarized light. 
Cherepkov and colleagues have also calculated spin polarization in various atoms shedding light on  dynamical origins of coefficients $\{A, \gamma\}$  and observed excellent agreement of their calculations with the experiments (see book chapter for pertinent references \cite{book1983advances}). 

While this framework is formally complete and provides a kinematic picture of spin-resolved photoionization, the dynamical properties and origin of the enantio-sensitive remain hidden. An intuitive picture for the enantio-sensitive parameters would be to express the parameters $\{D,C,B_1,B_2,B_3\}$ as pseudoscalars that change change sign upon switching enantiomers. For the case of PECD, the expression derived by Ritchie \cite{ritchie1976theory} can be equivalently written as\footnote{Here, all quantities are defined in the molecular frame.} 
\begin{align}
b_{1,0} \propto \int d\Theta_k \left[ \fullvec{k} \cdot \left( i \fullvec{D}_{\fullvec{k}}^* \times \fullvec{D}_{\fullvec{k}} \right) \right]
\end{align}
which allows for a physically transparent picture regarding the strength of the PECD signal \cite{ordonez2018generalized}, i.e., the strength of PECD is proportional to the flux of the geometric propensity field through the surface of the energy shell. This now raises the question on whether a similar intuitive procedure can be done for Eq. \eqref{eq:parameters} which we address in the next section.  

\section{Vectorial formulation of the photoionization parameters}
\label{sec:vector-form}

Let us now adopt the vectorial formulation \cite{ordonez2022geometric,ordonez2018generalized} into Cherepkov's approach \cite{cherepkov1983manifestations}. Here, we will show that the dynamical origin of spin- and enantio-sensitive photoionization observables arise from two intrinsic spin-resolved quantities which are quantified by the spin-resolved transition dipoles $\fullvec{D}_{\fullvec{k}^M,\mu^M}^M$:
\begin{subequations}
\begin{align}
\left( \vec{\mathbb{B}}_{\fullvec{k}^M}^M \right)_{\mu_1^M,\mu_2^M} = i \fullvec{D}_{\fullvec{k}^M,\mu_1^M}^{M*} \times \fullvec{D}_{\fullvec{k}^M,\mu_2^M}^M,
\label{eq:SR-Bfield}
\end{align}
\begin{align}
\fullvec{S}_{\fullvec{k}^M}^M = \sum_{\mu_1^M,\mu_2^M} \left( \fullvec{D}_{\fullvec{k}^M,\mu_1^M}^{M*} \cdot \fullvec{D}_{\fullvec{k}^M,\mu_2^M}^M \right) \unitvec{\sigma}_{\mu_2^M,\mu_1^M}
\end{align}
and an extrinsic mechanism which is the directional bias that is introduced by the well-defined direction of light polarization
\begin{align}
\left( \vec{\mathbb{K}}_{\fullvec{k}^M}^M \right)_{\mu_1^M,\mu_2^M} =  \fullvec{D}_{\fullvec{k}^M,\mu_1^M}^{M*} \left( \unitvec{k}^M \cdot \fullvec{D}_{\fullvec{k}^M,\mu_2^M}^M  \right).
\end{align}
\end{subequations}
Consequently, this means that the photoionization parameters, Eq. \eqref{eq:parameters}, can be expressed as moments of $\vec{\mathbb{B}}_{\fullvec{k}^M}^M$, $\fullvec{S}_{\fullvec{k}^M}^M$, and $ \vec{\mathbb{K}}_{\fullvec{k}^M}^M$ with respect to the momentum $\unitvec{k}^M$ and spin operator $\unitvec{\sigma}^M$. This effectively reduces the number of relevant quantities from ten to only three, which allows us to shed light on the dynamical origins of these parameters. The full details of the succeeding calculations are presented in Appendix \ref{app:vectors}.

It will be convenient to only rotate the photoelectron's spin function from the laboratory to molecular frame, 
\begin{align}
	\ket{\chi_{\mu^L}} =\sum_{\mu^M}\mathscr{D}_{\mu^L,\mu^M}^{1/2}|\chi_{\mu^M}\rangle
\end{align}
where, 
\begin{equation}
	\mathscr{D}^{1/2} = 
	\begin{bmatrix}
		e^{-i(\alpha+\gamma)/2}\cos\left(\frac{\beta}{2}\right) & -e^{-i(\alpha-\gamma)/2}\sin\left(\frac{\beta}{2}\right)\\
		e^{i(\alpha-\gamma)/2}\sin\left(\frac{\beta}{2}\right) & e^{i(\alpha+\gamma)/2}\cos\left(\frac{\beta}{2}\right)
	\end{bmatrix}.
    \label{eq:wigner-1/2}
\end{equation} 	
Next, we express the spin projection operator $\mathbb{\unitvec{P}}_{\unitvec{s}^{L}}$ in terms of the spherical harmonics $Y_{\ell_s,m_s}(\unitvec{s}^{L})$, such that,
\begin{align}
	\mathbb{\unitvec{P}}_{\unitvec{s}^{L}} =&\dfrac{1}{2}
	\begin{bmatrix}
		1+\sqrt{\dfrac{4\pi}{3}}Y_{1,0}(\unitvec{s}^{L}) & \sqrt{\dfrac{8\pi}{3}}Y_{1,-1}(\unitvec{s}^{L})\\
		-\sqrt{\dfrac{8\pi}{3}}Y_{1,1}(\unitvec{s}^{L}) & 1-\sqrt{\dfrac{4\pi}{3}}Y_{1,0}(\unitvec{s}^{L})
	\end{bmatrix}.
    \label{eq:projectorYs}
\end{align}
In this representation, the photoionization matrix element is now 
\begin{align}
	\langle \psi_o |  &(\fullvec{d}^L \cdot \fullvec{E}^L)^* |  I \Psi_{I,\fullvec{k}^M,\mu^L}^{(-)} \rangle \nonumber \\
	=& \langle \psi_o |  (\fullvec{d}^L \cdot \fullvec{E}^L)^* |  I \Psi_{I,\fullvec{k}^M,\mu^L}^{(-)} \rangle \nonumber \\
	%
	=&  \sum_{\mu^M } \mathscr{D}^{1/2}_{\mu^L,\mu^M}   (\fullvec{D}^{L*}_{\fullvec{k}^M,\mu^M}\cdot\fullvec{E}^{L*}) 
		\label{eq:dme}
\end{align}
where, $\fullvec{D}^{L}_{\fullvec{k}^M,\mu^M} \equiv \langle I \Psi_{I,\fullvec{k}^M,\mu^M}^{(-)} | \fullvec{d}^L | \psi_o \rangle$. For brevity, we will drop the index $I$ that denotes the ion channel. 

Repeating the same steps in which the photoionization rate for a given molecular orientation is obtained by taking the average of the spin projection operator, we get  
\begin{align}
W^L(\unitvec{k}^L,\unitvec{s}^L,\rho)  =& \sum_{\ell_s,m_s}  \mathcal{W}_{\ell_s,m_s}^L(\unitvec{k}^L,\rho)Y_{\ell_s,m_s}(\unitvec{s}^L)
\label{eq:yield0}
\end{align}
where, $\mathcal{W}_{\ell_s,m_s}^L(\unitvec{k}^L,\rho)$ is the `spin-weighted' yield:
\begin{subequations}
\begin{align}
\mathcal{W}_{\ell_s,m_s}^L & (\unitvec{k}^L,\rho) \nonumber \\
=& \sum_{\mu_1^M,\mu_2^M} (-1)^{m_s} \sqrt{\dfrac{\pi}{2\ell_s+1}} \, \Lambda^{(\ell_s,m_s)}_{\mu_2^M,\mu_1^M}  \nonumber \\
&\times \left( \fullvec{D}_{\fullvec{k}^M,\mu_1^M}^{L*}\cdot\fullvec{E}^{L*} \right) \left( \fullvec{D}_{\fullvec{k}^M,\mu_2^M}^{L}\cdot\fullvec{E}^{L} \right)
\label{eq:yield}
\end{align}
\begin{align}
	\Lambda^{(\ell_s,m_s)}_{\mu_2^M,\mu_1^M} = 
	\begin{cases}
		\delta_{\mu_1^M,\mu_2^M} \quad ; & \ell_s = 0 \\
		\left( \unitvec{\sigma}_{\mu_2^M,\mu_1^M}^L\cdot\unitvec{\epsilon}_{-m_s}^L \right)  \quad ;& \ell_s =1 
	\end{cases}
\end{align}
\begin{equation}
	\unitvec{\sigma}_{\mu_2^M, \mu_1^M}^L\equiv\langle\chi_{\mu_2^M}|\unitvec{\sigma}^L|\chi_{\mu_1^M}\rangle, 
\end{equation}
\begin{equation}
	\unitvec{\epsilon}_{0}^L = \unitvec{z}^L \quad , \quad \unitvec{\epsilon}_{\pm1}^L = \dfrac{\mp (\unitvec{x}^L \pm i \unitvec{y}^L)}{\sqrt{2}}.
\end{equation}
\end{subequations}
The momentum and spin resolved photoionization yield is finally obtained by taking the average over all molecular orientations,  
\begin{align}
	W^L(\unitvec{k}^L,\unitvec{s}^L) =& \sum_{\ell_s,m_s}\int d\rho \mathcal{W}_{\ell_s,m_s}^L(\unitvec{k}^L,\rho) Y_{\ell_s,m_s}(\unitvec{s}^L) 
	\label{eq:yield2}
\end{align}
The first term of Eq. \eqref{eq:yield2}, $\ell_s=m_s=0$, is the usual momentum-resolved photoionization yield and is not sensitive to spin measurement since it takes the sum of the contributions from spin-up and down photoelectrons. The effects of taking the photoelectron spin into consideration is now encapsulated in the terms $\ell_s=1$.

Since $W^{L}(\unitvec{k}^{L},\unitvec{s}^{L})$ depends on two directions, we can equivalently perform a multipolar expansion 
\begin{align}
	W^{L}(\unitvec{k}^{L},\unitvec{s}^{L}) =\sum b_{\ell,m_{\ell},\ell_{s},m_{s}}Y_{\ell,m_{\ell}}(\unitvec{k}^{L})Y_{\ell_{s},m_{s}}(\unitvec{s}^{L}), 
	\label{eq:yield1}
\end{align}
wherein, the coefficients $b_{\ell,m_{\ell},\ell_{s},m_{s}}$ now encode any information about the molecule and ionizing field. Comparing Eqs. \eqref{eq:yield0} and \eqref{eq:yield1} we see that 
\begin{align}
	b_{\ell,m_{\ell},\ell_{s},m_{s}} =& \int d\Theta_k^L Y_{\ell,m_{\ell}}^*(\unitvec{k}^{L}) \int d\rho \mathcal{W}_{\ell_s,m_s}^L(\unitvec{k}^L,\rho),
	\label{eq:coeff_b0}
\end{align}
where, $\int d\Theta_k^L$ is the integral over all photoelectron directions.
Using the definition of a rotated function, $\mathcal{W}_{\ell_s,m_s}^L(\unitvec{k}^L,\rho) =  \mathcal{W}_{\ell_s,m_s}^M(\unitvec{k}^M,\rho)$, we can equivalently write $\mathcal{W}_{\ell_s,m_s}^M(R_\rho^{-1}\unitvec{k}^L,\rho)=\mathcal{W}_{\ell_s,m_s}^L(\unitvec{k}^L,\rho)$, wherein, $R_\rho$ is the Euler rotation matrix that takes vectors from the molecular frame to the lab frame\footnote{This follows from the definition of a rotated function \cite{brink1968angular}, wherein $W^L(\unitvec{k}^L,\hat{s}^L,\rho)=W^M(\unitvec{k}^M,\hat{s}^M,\rho)$. Moreover, note that $\mu^M$ refers to either spin up or down in z-axis of the molecular frame $\unitvec{\zeta}^M$. Thus, to rotate $\unitvec{\sigma}_{\mu_2^M,\mu_1^M}^M$ from the molecular frame to the laboratory frame we have $\unitvec{\sigma}_{\mu_2^M,\mu_1^M}^L=R_\rho\unitvec{\sigma}_{\mu_2^M,\mu_1^M}^M$.}. Thus, Eq. \eqref{eq:coeff_b0} now takes the form 
\begin{align}
b_{\ell,m_{\ell},\ell_{s},m_{s}} =& \int d\rho \int d\Theta_k^L Y_{\ell,m_{\ell}}^*(\unitvec{k}^{L})  \mathcal{W}_{\ell_s,m_s}^M(R_\rho^{-1}\unitvec{k}^L,\rho) \nonumber \\
=& \int d\Theta_k^M \int d\rho Y_{\ell,m_{\ell}}^*(R_\rho\unitvec{k}^{M}) \mathcal{W}_{\ell_s,m_s}^M(\unitvec{k}^M,\rho),
\label{eq:coeff_b}
\end{align}
which follows from the change of variable $\unitvec{k}^L=R_\rho\unitvec{k}^M$. 

Note that when written in component form, the transition dipole $\fullvec{D}_{\fullvec{k}^M,\mu^M}^L$ and $\unitvec{\sigma}^L_{\mu^M,\nu^M}$ both transform as rank-1 tensors according to the Wigner matrix $\mathscr{D}^{(1)}(\rho)$, while the spherical harmonic $Y_{\ell,m_\ell}(\unitvec{k}^L)$ transforms as a rank-$\ell$ tensor according to $\mathscr{D}^{(\ell)}(\rho)$. It then follows from the orthogonality relations of the Wigner matrices that $b_{\ell,m_\ell,0,0}=0$ when $\ell>2$ while $b_{\ell,m_\ell,1,m_s}=0$ when $\ell>3$. For circularly polarized light, explicit evaluation of Eq. \eqref{eq:coeff_b} will lead to thirteen non-vanishing $b_{\ell,m_\ell,\ell_s,m_s}$ since the cylindrical symmetry of Eq. \eqref{eq:yield1} around $\unitvec{z}^L$ constrains $b_{\ell,m_\ell,\ell_s,m_s}$ to vanish when $m_\ell+m_s\neq0$, i.e., the non-vanishing coefficients are $\{b_{\ell\leq2,0,0,0}, b_{\ell\leq3,m_\ell,1,-m_\ell}\}$. 

To evaluate the orientation averaging $\int d\rho$ in Eq. \eqref{eq:coeff_b}, note that the spherical harmonics $Y_{\ell,m_\ell}(\unitvec{k}^L)$ are simply polynomials of $\{k_x,k_y,k_z\}$ such that the relevant integrals will have the general form 
\begin{align}
I = \int d\rho & \, \Lambda^{(\ell_s,m_s)}_{\mu_2^M,\mu_1^M} \left( \fullvec{D}_{\fullvec{k}^M,\mu_1^M}^{L*}\cdot\fullvec{E}^{L*} \right) \left( \fullvec{D}_{\fullvec{k}^M,\mu_2^M}^{L}\cdot\fullvec{E}^{L} \right)  \nonumber \\
& \times (\unitvec{k}^L \cdot \unitvec{x}^L)^p  (\unitvec{k}^L \cdot \unitvec{y}^L)^q(\unitvec{k}^L \cdot \unitvec{z}^L)^r
\label{eq:coeff_ave}
\end{align} 	
where, $p+q+r\leq \ell$. The vectors in Eq. \eqref{eq:coeff_ave} can be grouped into two sets, i.e., $\{\unitvec{x}^L,\unitvec{y}^L,\unitvec{z}^L,\unitvec{\epsilon}_{m_s}^L,\fullvec{E}^L\}$ which are fixed in the laboratory frame, and $\{\unitvec{k}^M,\fullvec{D}^M_{\fullvec{k}^M,\mu^M},\unitvec{\sigma}_{\mu_2^M,\mu_1^M}^M\}$ which are fixed in the molecular frame but appear above to be rotated into the laboratory frame $(\fullvec{a}^L=R_\rho\fullvec{a}^M)$. This now allows us to use the technique in Ref. \cite{andrews1977three} such that evaluating Eq. \eqref{eq:coeff_ave} will have the general form $I=\sum_{ij}g_i M_{ij}f_j$, where, $g_i$ and $f_j$ are rotational invariants that are formed from the set of vectors fixed in the molecular and laboratory frame, respectively, while $M_{ij}$ are coupling constants between the molecular and laboratory rotational invariants. 

Performing the necessary operations, Eqs. \eqref{eq:cherepkov-sigma}-\eqref{eq:cherepkov-B3} can now be expressed (see Appendix \ref{app:recover} for the proof) in a compact form that highlights their dynamical origin\footnote{Here, we omit the superscript $M$ since all quantities are defined in the molecular frame.}:
\begin{widetext}
\begin{subequations}\label{eq:dynamic}
\begin{minipage}{0.38\textwidth}
\begin{align}
\sigma_{\text{cross}} =& \frac{1}{3} S_0 \left| \fullvec{E}^L \right|^2
\label{eq:para-sigma-gen}
\end{align}
\begin{align}
\beta =& -\dfrac{1}{2} + \frac{3}{2S_0}  \int d\Theta_k \, \text{Tr} \left( \unitvec{k} \cdot \vec{\mathbb{K}}_{\fullvec{k}} \right) 
\label{eq:para-beta-gen} 
\end{align}    
\begin{equation}
A = \dfrac{1}{2 S_0} \int d\Theta_k \, \text{Tr} \left( \unitvec{\sigma} \cdot \vec{\mathbb{B}}_{\fullvec{k}} \right)
\label{eq:para-A-gen}
\end{equation}
\begin{align}
\eta=&\frac{3}{2S_0}  \int d\Theta_k \, \text{Re}\left[ \unitvec{k} \cdot \text{Tr} \left( \unitvec{\sigma} \times \vec{\mathbb{K}}_{\fullvec{k}} \right) \right] 
\label{eq:para-eta-gen}
\end{align}
\end{minipage}
\begin{minipage}{0.6\textwidth}
\begin{align}
\gamma = \frac{1}{2S_o} \int d\Theta_k \, & \text{Tr}  \left[ \left( \unitvec{k} \times \unitvec{\sigma} \right) \cdot \left( \unitvec{k} \times \vec{\mathbb{B}}_{\fullvec{k}} \right)   -  2 \left( \unitvec{k} \cdot \unitvec{\sigma} \right) \left( \unitvec{k} \cdot \vec{\mathbb{B}}_{\fullvec{k}} \right) \right]
\label{eq:para-gamma-gen}    
\end{align}
\begin{equation}
D = \dfrac{3}{2 S_0} \int d\Theta_k \left[ \unitvec{k} \cdot \text{Tr} \left( \vec{\mathbb{B}}_{\fullvec{k}}  \right) \right]
\label{eq:para-D-gen}
\end{equation}
\begin{equation}
C = \dfrac{3}{4 S_0} \int d\Theta_k \left[ \unitvec{k} \cdot \text{Tr} \left(  \unitvec{\sigma} \times \vec{\mathbb{B}}_{\fullvec{k}} \right) \right]
\label{eq:para-C-gen}
\end{equation}
\begin{align}
B_1 
%
%
=& \dfrac{3}{4S_0} \int d\Theta_k \,  \left\{ \left( \unitvec{k} \cdot \fullvec{S}_{\fullvec{k}}  \right) +  \text{Tr}\left[  \left(\unitvec{k}\cdot\unitvec{\sigma}\right) \left( \unitvec{k}\cdot\vec{\mathbb{K}}_{\fullvec{k}} \right) \right] \right\}
\label{eq:para-B1-gen}
\end{align}
\end{minipage}
\begin{align}
B_2 
=&  -\dfrac{3}{2S_0}  \int d\Theta_k \, \text{Tr} \left\{ \text{Re}\left[ \left( \unitvec{k} \times \unitvec{\sigma} \right) \cdot \left( \unitvec{k} \times \vec{\mathbb{K}} \right) \right] \right\}
\label{eq:para-B2-gen}    
\end{align}
\begin{align}
B_3 
=&  \dfrac{3}{4S_0}  \int d\Theta_k \, \left\{ \left( \unitvec{k} \cdot \fullvec{S}_{\fullvec{k}}  \right) +  \text{Tr} \left\{ 2 \text{Re}\left[ \left( \unitvec{k} \times \unitvec{\sigma} \right) \cdot \left( \unitvec{k} \times \vec{\mathbb{K}} \right) \right] -3 \left(\unitvec{k}\cdot\unitvec{\sigma}\right) \left( \unitvec{k}\cdot\vec{\mathbb{K}}_{\fullvec{k}} \right) \right\} \right\}
\label{eq:para-B3-gen}    
\end{align}
\end{subequations}
\end{widetext}
where, the trace is performed in spin-space, and $S_0$ denotes the total yield:
\begin{align}
	S_0 =  \sum_{\mu} \int d\Theta_k \, \left| \fullvec{D}_{\fullvec{k},\mu} \right|^2
\end{align}

\begin{figure}[t!]
	\centering
	\includegraphics[width=0.48\textwidth]{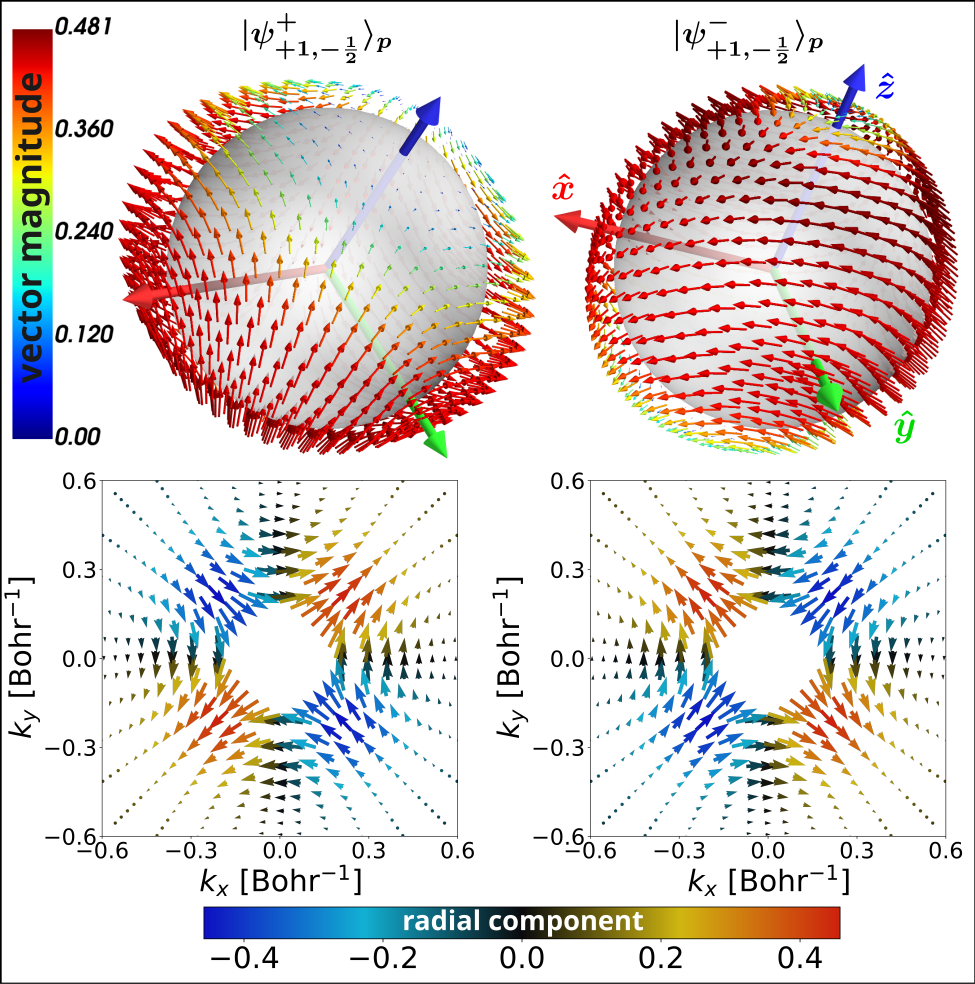}
	\caption{The momentum-resolved propensity field $\fullvec{\mathcal{B}}_{\fullvec{k}}$, Eq. \eqref{eq:D-flux}, in k-space for the chiral state $|\psi_{1,-\frac{1}{2}}^\pm\rangle_p$. The spheres correspond to $\fullvec{\mathcal{B}}_{\fullvec{k}}$ with $|\fullvec{k}|=0.525\,\text{Bohr}^{-1}$ where the vectors are colored as $|\fullvec{\mathcal{B}}_{\fullvec{k}}|$. The two-dimensional vector fields represent an equatorial cut in the $k_x-k_y$ plane and colored according to its radial component $(\unitvec{k}\cdot\fullvec{\mathcal{B}}_{\fullvec{k}})$. It can be seen that the propensity field for opposite enantiomers $R$ and $S$ are related via inversion $\fullvec{\mathcal{B}}_{\fullvec{k}}^{(R)}=\fullvec{\mathcal{B}}_{-\fullvec{k}}^{(S)}$.} 
	\label{fig:Bfield-texture}
\end{figure}

The standard momentum-resolved photoioniation of chiral molecules via circularly polarized light are described by the cross-section $\sigma_{\text{cross}}$, angular asymmetry parameter $\beta$, and PECD $D$. The inclusion of the photoelectron spin gives seven additional parameters $\{A,\eta,\gamma,C,B_1,B_2,B_3\}$ which are all proportional to the spin operator $\unitvec{\sigma}$.

The parameters, Eq. \eqref{eq:dynamic}, are driven by $\vec{\mathbb{B}}_{\fullvec{k}}$, $\fullvec{S}_{\fullvec{k}}$, and $\vec{\mathbb{K}}_{\fullvec{k}}$, however, notice that $\vec{\mathbb{B}}_{\fullvec{k}}$ does not appear with either $\fullvec{S}_{\fullvec{k}}$ or $\vec{\mathbb{K}}_{\fullvec{k}}$. Additionally, $\vec{\mathbb{B}}_{\fullvec{k}}$ drives the parameters $\{ A, \gamma, D, C \}$ which are the dicrhoic parts of the yield $W^L(\unitvec{k}^L,\unitvec{s}^L)$, while $\fullvec{S}_{\fullvec{k}}$ and $\vec{\mathbb{K}}_{\fullvec{k}}$ drives the non-dicrhoic part [see Eq. \eqref{eq:kinematic}]. Thus, $\vec{\mathbb{B}}_{\fullvec{k}}$ will only appear for light fields with a fixed polarization plane, e.g. circular and elliptically polarized light, while $\fullvec{S}_{\fullvec{k}}$ and $\vec{\mathbb{K}}_{\fullvec{k}}$ for arbitrary light fields. The enantio-sensitive parameters $\{ D, C, B_1, B_2, B_3 \}$ can be rewritten as the flux of an effective pseudovector involving $\{\vec{\mathbb{B}}_{\fullvec{k}}, \fullvec{S}_{\fullvec{k}},\vec{\mathbb{K}}_{\fullvec{k}}\}$, and using the relations Eq. \eqref{eq:relations} this flux is a pseudoscalar that changes sign upon switching enantiomer.

Equation \eqref{eq:dynamic} now makes the enantio-sensitivity of the parameters $\{ D, C, B_1, B_2, B_3 \}$ physically transparent. For opposite enantiomers $R$ and $S$, the transition dipoles are related as $\fullvec{D}_{\fullvec{k},\mu}^{(R)}=-\fullvec{D}_{-\fullvec{k},\mu}^{(S)}$ 
which implies the following relations: 
\begin{subequations} \label{eq:relations}
\begin{align}
\vec{\mathbb{B}}_{\fullvec{k}}^{(R)} = \vec{\mathbb{B}}_{-\fullvec{k}}^{(S)}
\end{align}
\begin{align}
\fullvec{S}_{\fullvec{k}}^{(R)} = \fullvec{S}_{-\fullvec{k}}^{(S)}
\end{align}
\begin{align}
\vec{\mathbb{K}}_{\fullvec{k}}^{(R)} = -\vec{\mathbb{K}}_{-\fullvec{k}}^{(S)}
\end{align}
\end{subequations}
To illustrate, consider the parameter $C$ as follows
\begin{align}
C^{(R)} =& \dfrac{3}{4 S_0} \int d\Theta_k \left[ \unitvec{k} \cdot \text{Tr} \left(  \unitvec{\sigma} \times \vec{\mathbb{B}}_{-\fullvec{k}}^{(S)} \right) \right] \nonumber \\
=& -\dfrac{3}{4 S_0} \int d\Theta_k \left[ \unitvec{k} \cdot \text{Tr} \left(  \unitvec{\sigma} \times \vec{\mathbb{B}}_{\fullvec{k}}^{(S)} \right) \right] =-C^{(S)}
\end{align}
which exactly behaves as a pseudoscalar that changes sign upon changing enantiomer. Similarly, consider the non-enantiosensitive parameter $A$, 
\begin{align}
A^{(R)} = \dfrac{1}{2 S_0} \int d\Theta_k \, \text{Tr} \left( \unitvec{\sigma} \cdot \vec{\mathbb{B}}_{-\fullvec{k}}^{(S)} \right) = A^{(S)}
\end{align}
which behaves as a scalar that does not change sign upon changing enantiomer. 

\begin{figure}[t!]
	\centering
	\includegraphics[width=0.48\textwidth]{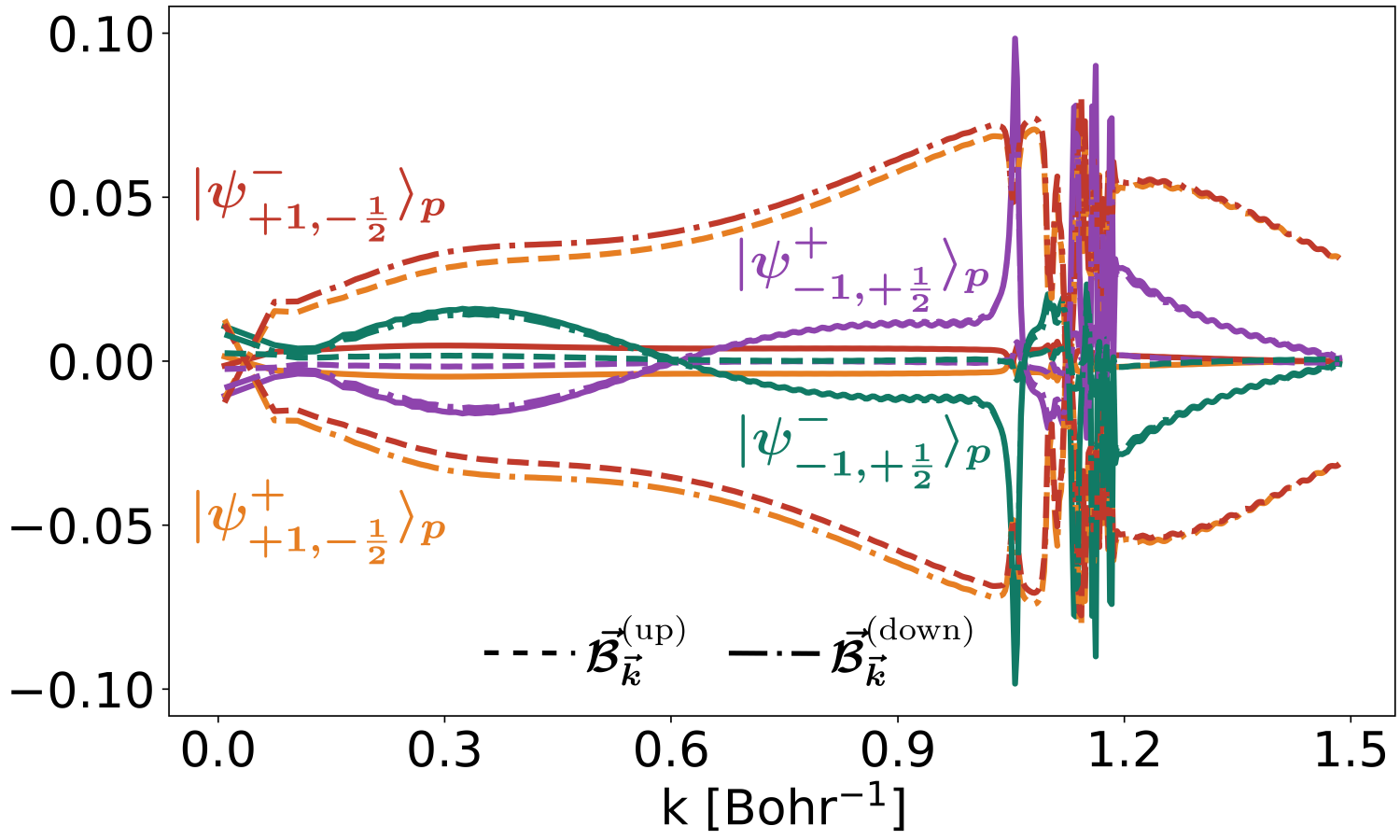}
	\caption{The coefficient $D$ (solid line) as well as the contributions of the spin-up (dashed line) and spin-down (dotted-dashed line), Eq. \eqref{eq:D-flux}, for the chiral states $|\psi_{1,-\frac{1}{2}}^\pm\rangle_p$ and $|\psi_{-1,\frac{1}{2}}^\pm\rangle_p$. The rapidly oscillating behavior of $D$ at higher values of \(k\) are due to the Fano resonances, leading up to the ionization threshold for the 3s electrons \cite{Samson2002,Carlstroem2024spinpolspectral}.} 
	\label{fig:coeffD}
\end{figure}

\section{Enantio-sensitive photoionization parameters}
\label{sec:physical_picture}

To quantify our results, we will use the same synthetic chiral argon system employed in Ref. \cite{flores2024_3} to provide a physical picture of the geometric quantitites $\{ \vec{\mathbb{B}}_{\fullvec{k}} , \fullvec{S}_{\fullvec{k}} , \vec{\mathbb{K}}_{\fullvec{k}} \}$ as well as quantify the parameters Eq. \eqref{eq:dynamic}. Synthetic chiral argon is constructed by combining excited-state orbitals, e.g.,
\begin{align}
	| \psi_{m_\ell,\mu}^{\pm} \rangle_p = \dfrac{1}{\sqrt{2}} \left( |4p_{m_\ell},\mu\rangle \pm |4d_{m_\ell},\mu\rangle \right).
	\label{eq:p_state}
\end{align}
which are inspired by analogous chiral hydrogenic states \cite{ordonez2019propensity}. Unlike hydrogen, the multielectron core potential in argon breaks inversion symmetry but the \emph{synthetic chirality} is stabilized by electron correlations.

\begin{figure}[t!]
	\centering
	\includegraphics[width=0.48\textwidth]{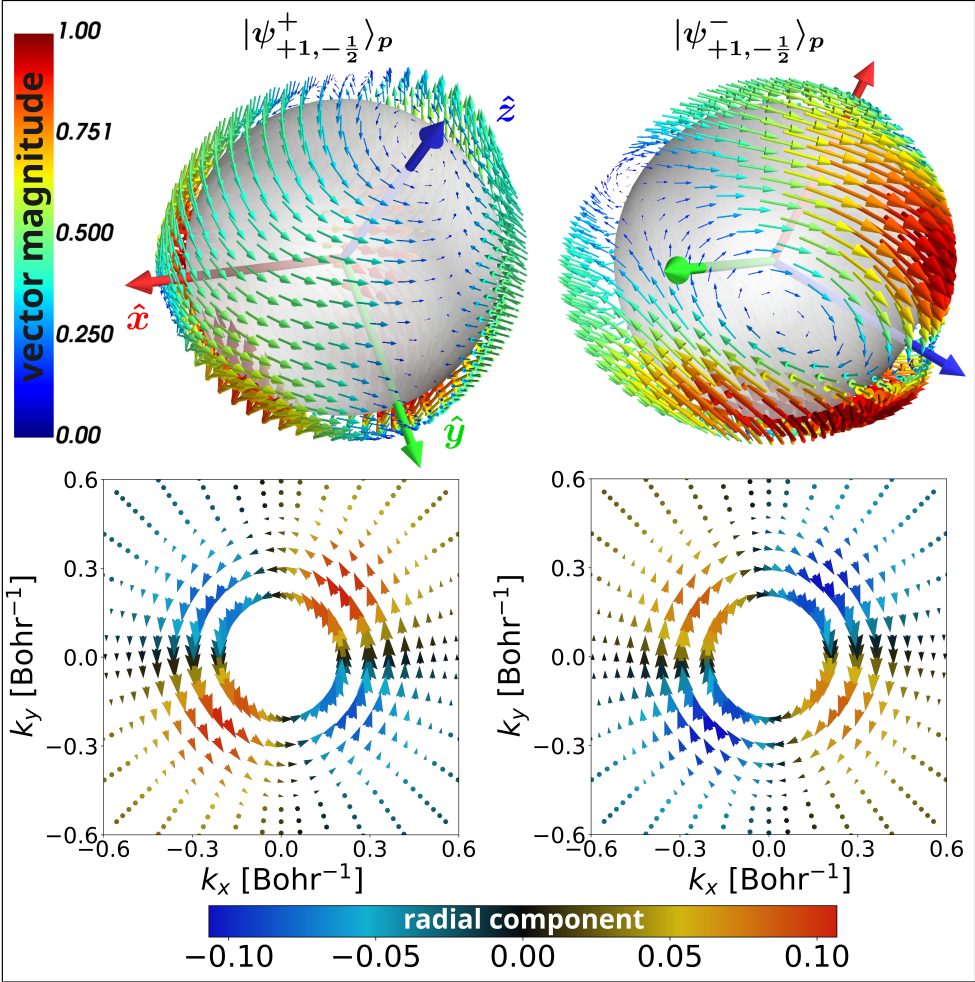}
	\caption{The momentum-resolved spin torque field $\fullvec{\tau}_{\fullvec{k}}$, Eq. \eqref{eq:C-flux}, in k-space for the chiral state $|\psi_{1,-\frac{1}{2}}^\pm\rangle_p$. The spheres correspond to $\fullvec{\tau}_{\fullvec{k}}$ with $|\fullvec{k}|=0.525\,\text{Bohr}^{-1}$ where the vectors are colored as $|\fullvec{\tau}_{\fullvec{k}}|$. The two-dimensional vector fields represent an equatorial cut in the $k_x-k_y$ plane and colored according to its radial component $(\unitvec{k}\cdot\fullvec{\tau}_{\fullvec{k}})$. It can be seen that the spin torque field for opposite enantiomers $R$ and $S$ are related via inversion $\fullvec{\tau}_{\fullvec{k}}^{(R)}=\fullvec{\tau}_{-\fullvec{k}}^{(S)}$.} 
	\label{fig:TorqueField-texture}
\end{figure}
    
\subsection{Propensity field}

The pseudovector $\vec{\mathbb{B}}_{\fullvec{k}}$ is referred to as the spin-resolved propensity field as it encodes the photoionization propensity rules. It is a matrix in spin space, and serves as the natural extension of the geometric propensity field for spinless systems, Eq. \eqref{eq:Bfield}. As can be seen from Eqs. \eqref{eq:kinematic} and \eqref{eq:dynamic}, the pseudovector $\vec{\mathbb{B}}_{\fullvec{k}}$ drives the dichroic and enantio-sensitive components of $W^L(\unitvec{k}^L,\unitvec{s}^L)$ described by the parameters $\{D,C\}$.

The parameter $D$, Eq. \eqref{eq:para-D-gen}, describes PECD which is equivalent to the flux of the spin-symmetric propensity field (see Fig. \ref{fig:Bfield-texture}) through the surface of a sphere with radius $k$:
\begin{subequations} \label{eq:D-flux}
\begin{align}
D = \dfrac{3}{2S_0} \int d\unitvec{\Theta}_k \cdot \fullvec{\mathcal{B}}_{\fullvec{k}}
\end{align}
\begin{align}
\fullvec{\mathcal{B}}_{\fullvec{k}} = \text{Tr}( \vec{\mathbb{B}}_{\fullvec{k}} ) 
= \left( \vec{\mathbb{B}}_{\fullvec{k}} \right)_{+\frac{1}{2},+\frac{1}{2}} + \left( \vec{\mathbb{B}}_{\fullvec{k}} \right)_{-\frac{1}{2},-\frac{1}{2}} 
\end{align}
\end{subequations}
where, $d\unitvec{\Theta}_k = d\Theta \unitvec{k}$. The expression Eq. \eqref{eq:D-flux} is equal to the PECD we derived in \cite{ordonez2018generalized}, as well as the original expression derived by Ritchie \cite{ritchie1976theory}. The strength of the PECD now encodes the interplay between the flux of the propensity field corresponding to the degenerate spin-up and spin-down photoionization channels. Figure \ref{fig:coeffD} shows a case when $D$ is negligible because the flux of the contributions from the spin-up and spin-down propensity field has opposite signs, and an alternative case when the flux of the spin-up propensity field is negligible.

\begin{figure}[t!]
	\centering
	\includegraphics[width=0.48\textwidth]{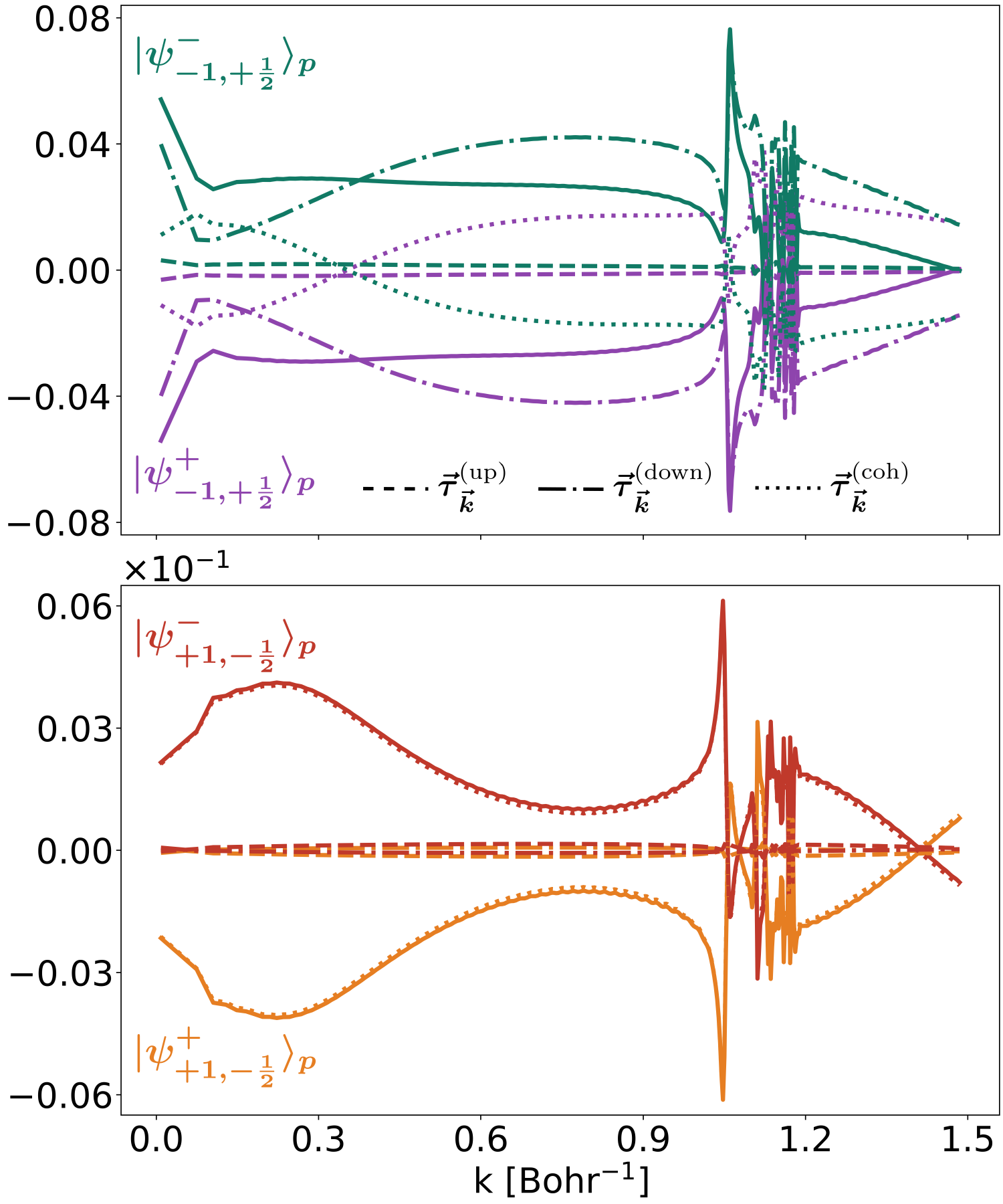}
	\caption{The coefficient $C$ (solid line) as well as the contributions of the spin-up (dashed line) and spin-down (dotted-dashed line), Eq. \eqref{eq:D-flux}, for the chiral states $|\psi_{1,-\frac{1}{2}}^\pm\rangle_p$ and $|\psi_{-1,\frac{1}{2}}^\pm\rangle_p$.} 
	\label{fig:coeffC}
\end{figure}

\begin{figure}[t!]
	\centering
	\includegraphics[width=0.48\textwidth]{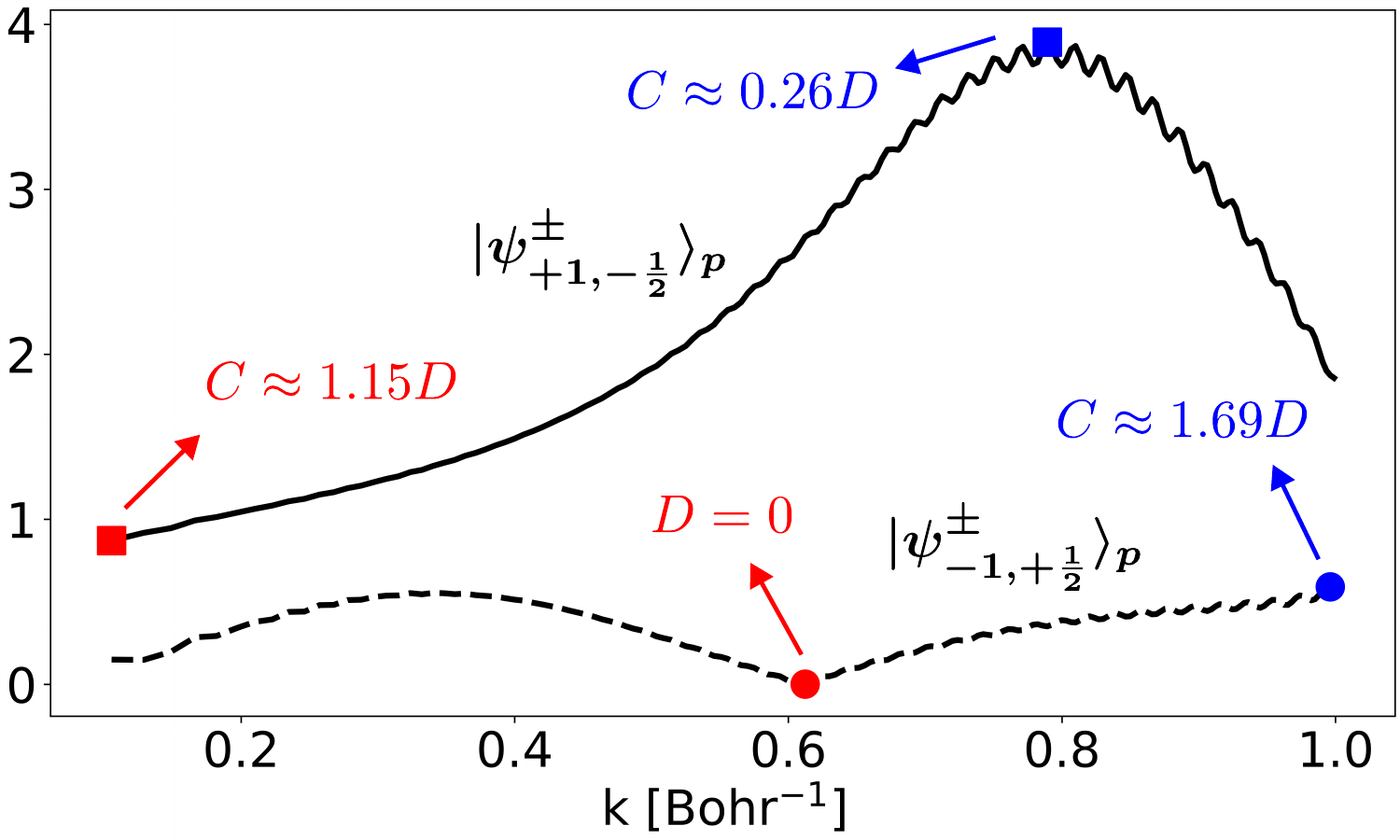}
	\caption{The ratio $|D/C|$ which shows the relative strength of the transversal spin-polarization driven by the parameter $C$ compared to the strength of PECD driven by $D$ for the states $|\psi_{1,-\frac{1}{2}}^\pm\rangle_p$ (solid line) and $|\psi_{-1,\frac{1}{2}}^\pm\rangle_p$ (dashed line) for $0.1\,\text{Bohr}^{-1}<k<1.0\,\text{Bohr}^{-1}$} 
	\label{fig:ratio-DC}
\end{figure}

The parameter $C$, Eq. \eqref{eq:para-C-gen}, is both spin- and enantio-sensitive, and arises from the flux of a spin-torque pseudovector [see Fig. \ref{fig:TorqueField-texture}]:
\begin{subequations} \label{eq:C-flux}
\begin{align}
C = \dfrac{3}{4S_0} \int d \unitvec{\Theta}_k \cdot \fullvec{\tau}_{\fullvec{k}}
\end{align}
\begin{align}
	\fullvec{\tau}_{\fullvec{k}} = \text{Tr} \left( \unitvec{\sigma} \times \vec{\mathbb{B}}_{\fullvec{k}} \right) = \fullvec{\tau}_{\fullvec{k}}^{(\text{up})} + \fullvec{\tau}_{\fullvec{k}}^{(\text{down})} + \fullvec{\tau}_{\fullvec{k}}^{(\text{coh})}, 
\end{align}
\end{subequations}
and describes how the molecular geometry imparts a torque on the photoelectron spin. The terms $\fullvec{\tau}_{\fullvec{k}}^{(\text{up})}$, $\fullvec{\tau}_{\fullvec{k}}^{(\text{down})}$, and $\fullvec{\tau}_{\fullvec{k}}^{(\text{coh})}$ arise from the spin-up, spin-down, and coherence terms of the propensity field $\vec{\mathbb{B}}_{\fullvec{k}}$. Alternatively, $C$ can be rewritten as 
\begin{subequations}
\begin{align}
C=\dfrac{3}{4S_0} \int d\Theta_k \, \text{Tr}\left[ \vec{\mathbb{B}}_{\fullvec{k}} \cdot \left( \unitvec{k} \times \nabla_k \mathbb{Y} \right) \right]
\end{align}
\begin{align}
\mathbb{Y} = \sqrt{\dfrac{4\pi}{3}} 
\begin{bmatrix}
			Y_{1,0}(\unitvec{k}) & \sqrt{2}Y_{1,-1}(\unitvec{k}) \\
			-\sqrt{2}Y_{1,1}(\unitvec{k}) & -Y_{1,0}(\unitvec{k})
		\end{bmatrix}
\end{align}
\end{subequations}
which shows that it is a multipole moment of the transversal part of the spin-resolved propensity field $\vec{\mathbb{B}}_{\fullvec{k}}$,

The contribution of $C$ to the yield $W^L(\unitvec{k}^L,\unitvec{s}^L)$ becomes maximal when the photoelectron momentum $\unitvec{k}^L$, spin-detection axis $\unitvec{s}^L$, and photon spin $\unitvec{\Xi}^L$ are mutually orthogonal [see Eq. \eqref{eq:kinematic}]. Therefore, $C$ drives a spin-polarization vortex in the polarization plane that rotates in opposite direction for opposite enantiomers, which is reminiscent of the Rashba effect in solids. Figure \ref{fig:coeffC} shows that the coherence between the spin-up and spin-down channels play a significant role on the strength of $C$. Figure \ref{fig:ratio-DC} shows the relative strength of the parameters $D$ and $C$ and that the transversal spin-polarization driven by $C$ is on the same order as PECD, and can even be significantly larger, i.e., $0.26D\leq C \leq 1.15D$ for the states $|\psi_{1,-\frac{1}{2}}^\pm\rangle_p$ and $C\geq1.69D$ for the states $|\psi_{-1,\frac{1}{2}}^\pm\rangle_p$.

\subsection{Bloch vector}

\begin{figure}[t!]
	\centering
	\includegraphics[width=0.48\textwidth]{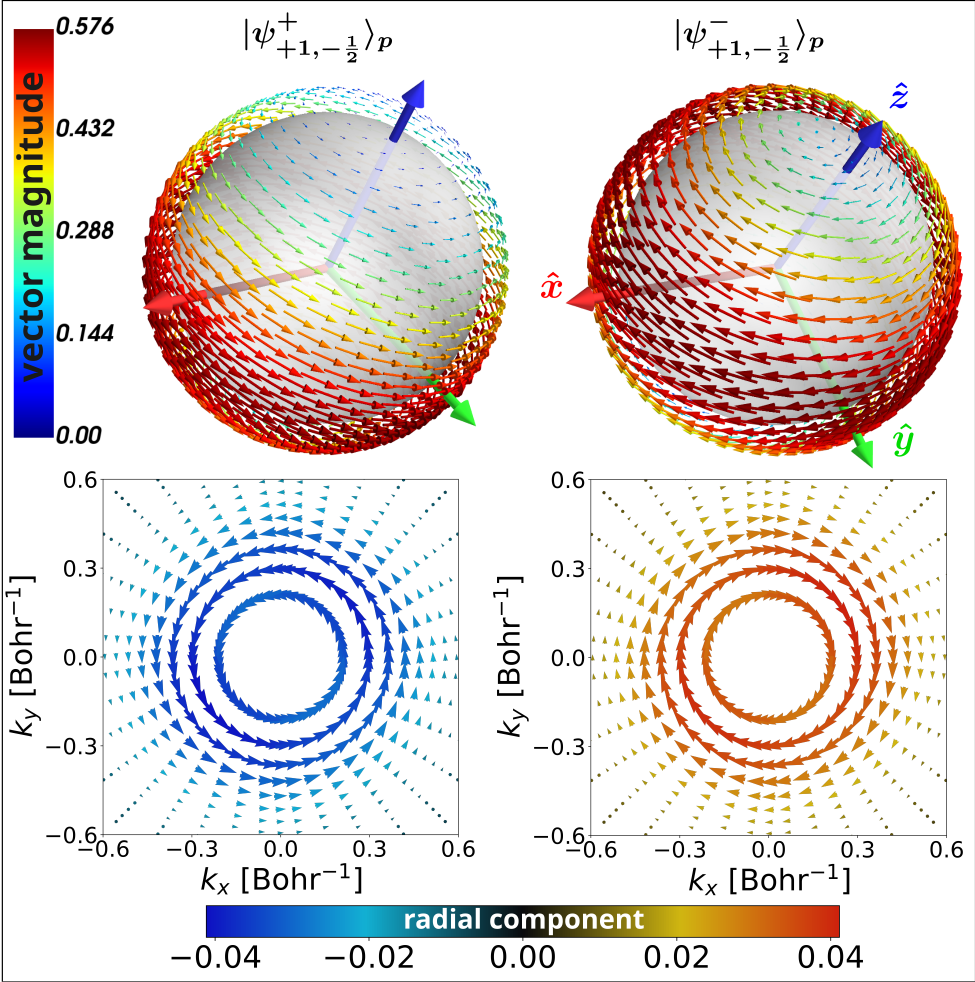}
	\caption{The momentum-resolved Bloch pseudovector $\fullvec{S}_{\fullvec{k}}$ in k-space for the chiral state $|\psi_{1,-\frac{1}{2}}^\pm\rangle_p$. The spheres correspond to $\fullvec{S}_{\fullvec{k}}$ with $|\fullvec{k}|=0.525\,\text{Bohr}^{-1}$ where the vectors are colored as $|\fullvec{S}_{\fullvec{k}}|$. The two-dimensional vector fields represent an equatorial cut in the $k_x-k_y$ plane and colored according to its radial component $(\unitvec{k}\cdot\fullvec{S}_{\fullvec{k}})$. It can be seen that the Bloch pseudovector field for opposite enantiomers $R$ and $S$ are related via inversion $\fullvec{S}_{\fullvec{k}}^{(R)}=\fullvec{S}_{-\fullvec{k}}^{(S)}$.} 
	\label{fig:BlochField-texture}
\end{figure}

The pseudovector $\fullvec{S}_{\fullvec{k}}$ is referred to as the momentum-resolved photoionization Bloch vector introduced in Ref. \cite{flores2024_3}, i.e., 
\begin{align}
\fullvec{S}_{\fullvec{k}^M}^M =& \sum_{\mu_1^M,\mu_2^M} \left( \fullvec{D}_{\fullvec{k}^M,\mu_1^M}^{M*} \cdot \fullvec{D}_{\fullvec{k}^M,\mu_2^M}^M \right) \unitvec{\sigma}_{\mu_2^M,\mu_1^M} \nonumber \\
=& \text{Tr}(\tilde{\varrho}\unitvec{\sigma}). 
\end{align}
where, $\tilde{\varrho}$ is the reduced density matrix of a two-level spin-1/2 system associated with a given photoelectron energy. Mathematically, it is obtained by performing a partial trace over the spatial contiuum states and degenerate ion channels, then performing orientation averaging over random molecular orientations. Figure \ref{fig:BlochField-texture} shows that the Bloch vector $\fullvec{S}_{\fullvec{k}}$ exhibits a vortex structure, thereby highlighting the importance of the coherence between the spin-up and spin-down channels. Meanwhile, we refer to the quantity $\vec{\mathbb{K}}_{\fullvec{k}}$ as an asymmetry pseudovector since it appears in the asymmetry parameter $\beta$ for momentum-resolved photoionization. 

As can be seen from Eqs. \eqref{eq:kinematic} and \eqref{eq:dynamic}, the pseudovectors $\fullvec{S}_{\fullvec{k}}$ and $\vec{\mathbb{K}}_{\fullvec{k}}$ drive the non-dichroic and enantio-sensitive components of $W^L(\unitvec{k}^L,\unitvec{s}^L)$ described by the parameters $\{B_1,B_2,B_3\}$. The parameters $B_1$ and $B_3$ can both be rewritten as the flux of an effective effective Bloch pseudovector $\fullvec{S}^{(\text{eff})}_{\fullvec{k}}$ involving $\fullvec{S}_{\fullvec{k}}$ and $\vec{\mathbb{K}}_{\fullvec{k}}$, while $B_2$ is purely the flux of the asymmetry pseudovecor, i.e., 
\begin{subequations} \label{eq:B1-flux}
\begin{align}
B_1 = \dfrac{3}{4S_0} \int d \unitvec{\Theta}_k \cdot \fullvec{S}^{(\text{eff},1)}_{\fullvec{k}}
\end{align}
\begin{align}
\fullvec{S}^{(\text{eff},1)}_{\fullvec{k}} = \fullvec{S}_{\fullvec{k}} + \text{Tr}\left[ \left( \unitvec{k}\cdot\vec{\mathbb{K}}_{\fullvec{k}} \right) \unitvec{\sigma} \right] 
\end{align}
\end{subequations}
\begin{subequations} \label{eq:B2-flux}
\begin{align}
B_2 = -\dfrac{3}{2S_0} \int d \unitvec{\Theta}_k \cdot \fullvec{\mathcal{K}}_{\fullvec{k}}
\end{align}
\begin{align}
\fullvec{\mathcal{K}}_{\fullvec{k}} = \text{Tr}\left\{ \text{Re}\left[ \unitvec{\sigma} \times \left( \unitvec{k} \times \vec{\mathbb{K}}_{\fullvec{k}} \right)  \right] \right\}
\end{align}
\end{subequations}
\begin{subequations} \label{eq:B3-flux}
\begin{align}
B_3 = \dfrac{3}{4S_0} \int d \unitvec{\Theta}_k \cdot \fullvec{S}^{(\text{eff},2)}_{\fullvec{k}}
\end{align}
\begin{align}
\fullvec{S}^{(\text{eff},2)}_{\fullvec{k}} = \fullvec{S}_{\fullvec{k}} + \text{Tr}\left\{ 2\text{Re}\left[ \unitvec{\sigma} \times \left( \unitvec{k} \times \vec{\mathbb{K}}_{\fullvec{k}} \right)  \right]  -3\left( \unitvec{k}\cdot\vec{\mathbb{K}}_{\fullvec{k}} \right) \unitvec{\sigma} \right\} 
\end{align}
\end{subequations}
These parameters are formed from the combinations of several $b_{\ell,m_\ell,\ell_s,m_s}$ coefficients [see Eq. \eqref{eq:parameters}] to obtain the vector structure in Eq. \eqref{eq:kinematic}, as such, these parameters will thereby need several detector setups, and clearly, are no longer linearly independent from each other. Specifically, we see from Eq. \eqref{eq:dynamic} that the parameter $B_2$ is contained in the parameter $B_3$.

We see from Eq. \eqref{eq:kinematic} that the parameter $B_1$ describes the helicity of the photoelectron since it projects the measured momentum $\unitvec{k}^L$ onto the spin-detection axis $\unitvec{s}^L$. The strength of this helicity signal is described by the flux of $\fullvec{S}^{(\text{eff},1)}_{\fullvec{k}}$ through the surface of the energy shell. Since this parameter is driven by the Bloch vector, then it will be non-zero even for the case of linearly polarized light. The parameter $B_1$ thereby presents a spin analog of photoelectron vortex dichroism (PEVD) \cite{planas2022ultrafast}, wherein the the photoelectron helicity encodes molecular chirality.

\begin{figure}[t!]
	\centering
	\includegraphics[width=0.48\textwidth]{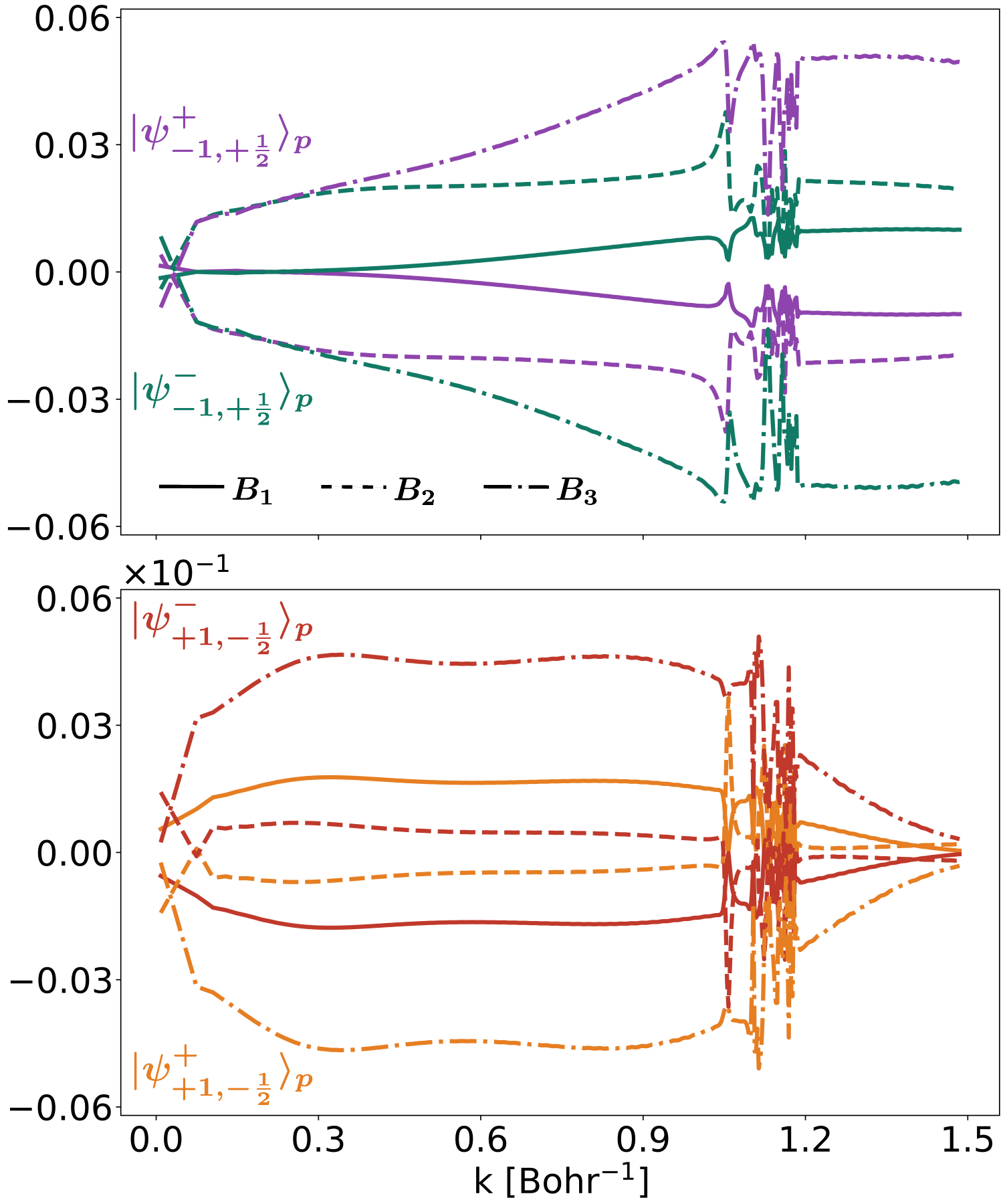}
	\caption{The coefficient $B_1$ (solid line), $B_2$ (dashed line), and $B_3$ (dotted-dashed line) for the chiral states $|\psi_{1,-\frac{1}{2}}^\pm\rangle_p$ and $|\psi_{-1,\frac{1}{2}}^\pm\rangle_p$. } 
	\label{fig:coeffB123}
\end{figure}

The contribution of the parameter $B_2$ to the yield $W(\unitvec{k}^L,\unitvec{s}^L)$ becomes maximal when the photoelectron momentum $\unitvec{k}^L$, spin-detection axis $\unitvec{s}^L$, and photon spin $\unitvec{\Xi}^L$ are collinear [see Eq. \eqref{eq:kinematic}]. Therefore, $B_2$ drives a non-dichroic PECD-like signal that is spin-polarized along the direction of photon spin. Last, the parameter $B_3$ is simply a higher pole observable. Figure \ref{fig:coeffB123} shows the value of the parameters $\{B_1,B_2,B_3\}$ and it can be seen these parameters is on the same order as PECD, and can also be significantly larger, thereby demonstrating that spin-resolved photoionization of chiral molecules presents alternative enantio-sensitive observables in the electric dipole approximation that can be larger than PECD.

\section{Spin polarization parameters}
\label{sec:total_sp}

As we have seen from Sec. \ref{sec:vector-form}, the inclusion of the photoelectron spin degree of freedom gives seven additional parameters $\{A,\eta,\gamma,C,B_1,B_2,B_3\}$ which are all proportional to the spin operator $\unitvec{\sigma}$. Among these, we have discussed the enantio-sensitive parameters $\{C,B_1,B_2,B_3\}$ in Sec. \ref{sec:physical_picture} which all arise from the expansion coefficients $A_{L,M_L,L_S,-M_L}$ wherein $L$ is odd, see Eq. \eqref{eq:parameters}. Similarly, the non-enantiosensitive parameters $\{A,\eta,\gamma\}$ all arise from the expansion coefficients $A_{L,M_L,L_S,-M_L}$ wherein $L$ is even. These symmetries of $A_{L,M_L,L_S,M_S}$ are consistent with recent findings of Ref. \cite{artemyev2026SP}.     

Among all the spin polarization parameters, of particular interest is the coefficient $A$, see Eq. \eqref{eq:para-A-gen}, which shows that the total spin polarization is non-enantiosensitive. Formally, this is obtained by taking the averaged value of $\unitvec{s}^L$, i.e., 
\begin{align}
\langle \unitvec{s}^L \rangle =& \dfrac{ \int d\rho \int d\Theta_k^L \int d\Theta_s^L W^L(\unitvec{k}^L,\unitvec{s}^L,\rho)\unitvec{s}^L}{\int d\rho \int d\Theta_k^L \int d\Theta_s^L W^L(\unitvec{k}^L,\unitvec{s}^L,\rho)} \nonumber \\
=& \left[ \dfrac{1}{3S_0} \int d\Theta_k^M \, \text{Tr} \left( \unitvec{\sigma}^M \cdot \vec{\mathbb{B}}_{\fullvec{k}^M}^M \right) \right] \unitvec{\Xi}^L \nonumber \\
=& \left( \dfrac{3}{2} A \right) \unitvec{\Xi}^L,
\label{eq:chi_gs}
\end{align}
which shows that the total spin polarizaon is along the direction of the photon spin. 

\begin{figure}[t!]
	\centering
	\includegraphics[width=0.48\textwidth]{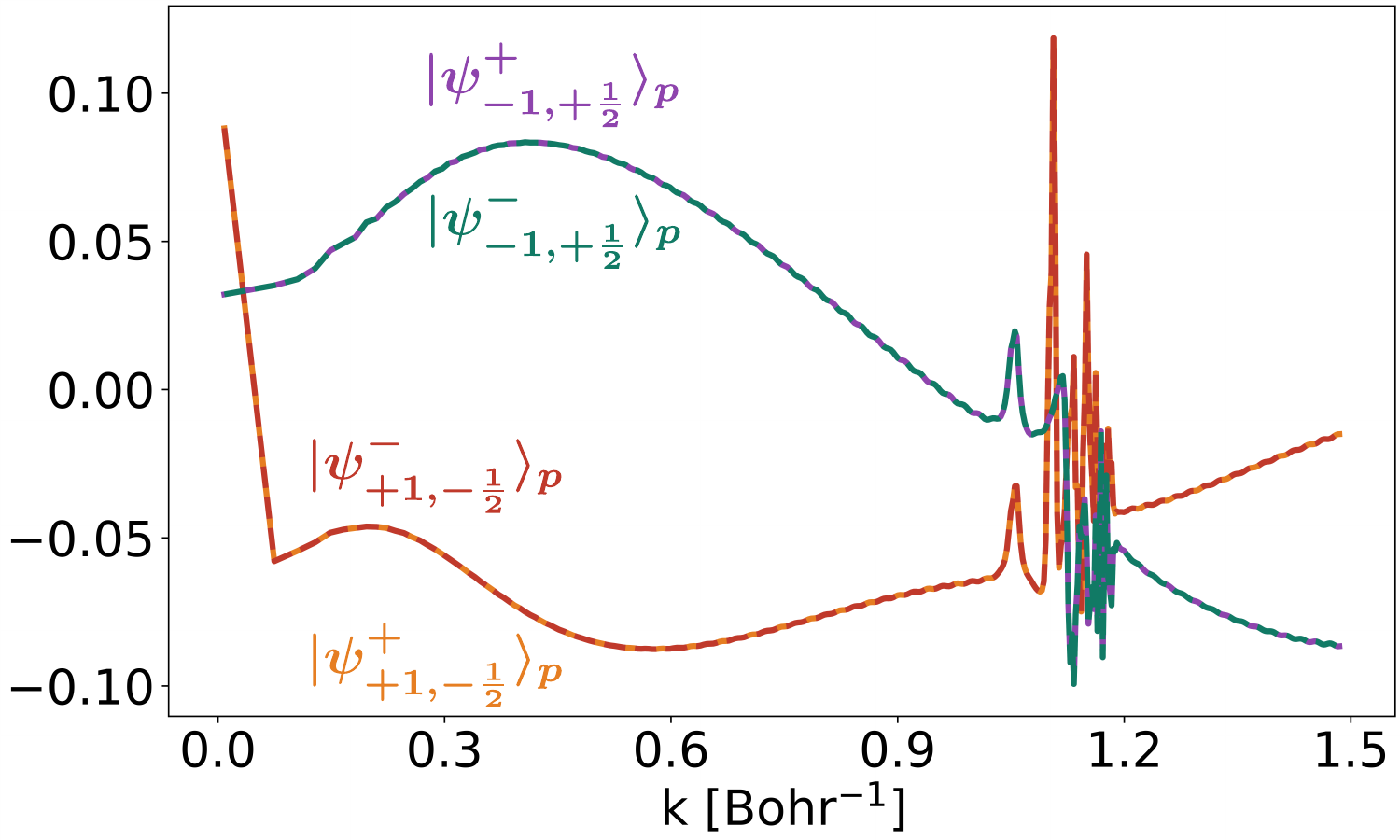}
	\caption{The geometric spin susceptibility $\chi_{\text{gs}}=(3/2)A$ for the chiral states $|\psi_{1,-\frac{1}{2}}^\pm\rangle_p$ and $|\psi_{-1,\frac{1}{2}}^\pm\rangle_p$. } 
	\label{fig:coeffA}
\end{figure}

The structure of Eq. \eqref{eq:chi_gs} is reminiscent to that of the standard magnetic susceptibility $\chi$, where the spin response to an external magnetic field $\fullvec{B}_{\text{ext}}$ is characterized by $\langle \unitvec{s} \rangle = \chi \fullvec{B}_{\text{ext}}$. Accordingly, the parameter $A$ may be interpreted as a \textit{geometric spin susceptibility}: 
\begin{equation}
\chi_{\text{gs}} = \dfrac{3}{2} A,
\end{equation}
a measure of how geometric structure mediates spin orientation in response to circularly polarized light. Figure \ref{fig:coeffA} shows the spin-polarization, $\chi_{\text{gs}}$ of the chiral states $|\psi_{1,-\frac{1}{2}}^\pm\rangle_p$ and $|\psi_{-1,\frac{1}{2}}^\pm\rangle_p$ which can reach up to $12\%$.

The form of the parameter $A$ as given in Eq. \eqref{eq:para-A-gen} exposes the geometric origin of spin polarization in photoionization: the spin-resolved propensity field associated with photoionization acts to orient the photoelectron spin. Moreover, Eq. \eqref{eq:chi_gs} represents an anomalous vectorial observable analogous to Class I observables introduced in Ref. \cite{ordonez2023geometric}. This observable has been experimentally detected the case of multiphoton ionization \cite{artemyev2024photoelectron} as well as atomic targets in the one-photon \cite{heinzmann1980experimental}.

\begin{figure}[t!]
	\centering
	\includegraphics[width=0.48\textwidth]{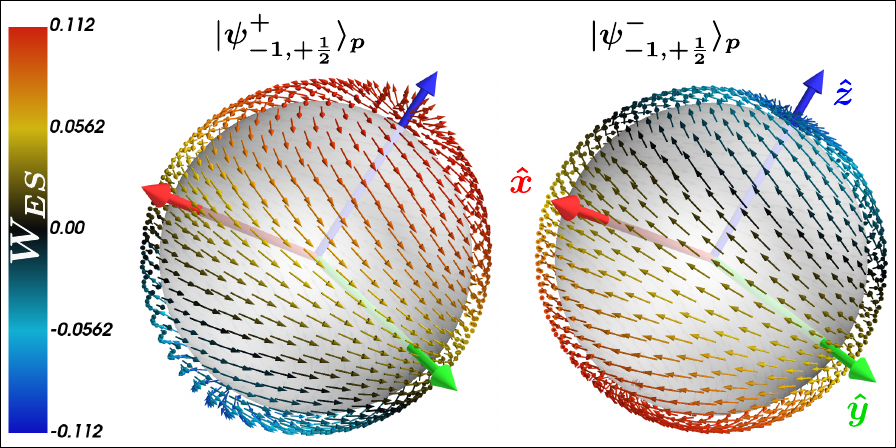}
	\caption{Spin texture that maximizes the enantio-sensitve part of the yield $W(\unitvec{k}^L,\unitvec{s}^L)$ for the chiral states $|\psi_{-1,\frac{1}{2}}^\pm\rangle_p$ at $k=1.055\,\text{Bohr}^{-1}$. The arrows are colored according to the enantiosensitive component of the yield $W_{ES}$. } 
	\label{fig:spin-texture}
\end{figure}

Although the total spin polarization, $A$, is not enantio-sensitive, the angular distribution of the spin polarization is enantio-sensitive which are encapsulated in the parameters $\{C,B_1,B_2,B_3\}$. Using Eq. \eqref{eq:kinematic}, we can construct a spin-texture that maximizes the enantio-sensitive part of the yield $W(\unitvec{k}^L,\unitvec{s}^L)$. This spin-texture defines the axis in which the spin-detector should be placed for a given photoelectron momentum $\fullvec{k}$ in order to obtain the maximal enantio-sensitive and spin-polarized yield $W_{ES}$. Indeed, Figure \ref{fig:spin-texture} shows that both $W_{ES}$ and the direction of $\unitvec{s}^L$ changes sign upon switching enantiomer. For the chosen chiral state $|\psi_{-1,\frac{1}{2}}^\pm\rangle_p$, we see that the spin-texture exhibits a vortex-like structure thereby indicating that the optimal detection scheme would be to place the spin-detector axis $\unitvec{s}^L$ perpendicular to the photoelectron momentum $\unitvec{k}^L$.

\section{Summary and Conclusion}
\label{sec:conc}

We have revisited the earlier of Cherepkov \cite{cherepkov1983manifestations} regarding spin-resolved one-photon ionization of chiral molecules, wherein it was shown that ten independent parameters are required to fully describe all spin- and momentum-resolved observables. Among these, five are enantio-sensitive. While the framework is formally complete and provides a kinematic picture of spin-resolved photoionization, the dynamical properties and origin of the enantio-sensitive parameters remain hidden. 

Here, we used an alternative vector approach \cite{ordonez2022geometric,ordonez2018generalized} and found that these ten parameters can all be expressed as multipole moments of the spin-resolved propensity field, $\fullvec{\mathbb{B}}_{\fullvec{k}}$, photoionization Bloch pseudovector, $\fullvec{S}_{\fullvec{k}}$, and asymmetry pseudovector $\fullvec{\mathbb{K}}_{\fullvec{k}}$, with respect to the momentum $\unitvec{k}$ and spin operator $\unitvec{\sigma}$. This effectively reduces the number of relevant quantitites from ten to three.

The strength of the enantio-sensitive parameters, $\{D,C,B_1,B_2,B_3\}$ are all described as the flux through the energy shell of an effective pseudovectors involving $\{ \fullvec{\mathbb{B}}_{\fullvec{k}},\fullvec{S}_{\fullvec{k}},\fullvec{\mathbb{K}}_{\fullvec{k}} \}$. The propensity field drives the dichroic parameters $D$ and $C$ which desribes PECD and a spin-polarization vortex in the light polarization plane which rotates in opposite directions for opposite enantiomers. The Bloch and asymmetry pseudovectors drive the non-dichroic parameters $\{B_1,B_2,B_3\}$ which describes a helicity signal, PECD-like signal that is spin polarized along the direction of photon spin, and a higher multipole observable, respectively.

To quantify the strength of these parameters, we used synthetic chiral argon as a toy model, and showed that the spin-polarized observables $\{C,B_1,B_2,B_3\}$ can be on the same order of magnitude as the PECD signal, and can even be significantly larger. This demonstrates that spin-resolved photoionization of chiral molecules provides more possible directions for chiral discrimination. 

Our results provide compact expressions for these observables which provide an intuitive picture on what determines the strength of these spin- and enantio-sensitive observables. The approach can be readily generalized to photoexcitation, multiphoton processes, and arbitrary field polarizations. Regardless of the specific driving conditions, the resulting spin- and enantio-sensitive observables are still controlled by the same three pseudovectors, underscoring their universal role as the primary generators of chirality-induced spin asymmetries, emphasizing their fundamental geometric origin and the universality of the mechanism identified here.


\section*{Acknowledgements}

O.S., A.F. O. and P.C.F. acknowledge ERC-2021-AdG project ULISSES, grant agreement No 101054696. Views and opinions expressed are however those of the author(s) only and do not necessarily reflect those of the European Union or the European Research Council. Neither the European Union nor the granting authority can be held responsible for them. A.F.O. acknowledges funding from the Royal Society URF/R1/201333, URF/ERE/210358, and URF/ERE/231177 and from the Deutsche Forschungsgemeinschaft (DFG, German Research Foundation) - 543760364. 



\begin{widetext}
\appendix

\section{Reconstruction of Cherpekov's vector structure for the momentum- and spin-resolved yield}
\label{app:cherepkov_yield}

For completeness, we present the construction of Eqs. \eqref{eq:kinematic}-\eqref{eq:parameters} based on the symmetry of the Wigner 3j-symbols in Eq. \eqref{eq:cherepkov_Y}, i.e., 
\begin{subequations} \label{eq-app:cherepkov_Y}
\begin{align}
W^{L}(\unitvec{k}^{L},\unitvec{s}^{L})= & \sum A_{L,M_{L},S,M_{S}}Y_{L,M_{L}}(\unitvec{k}^L)Y_{S,M_{S}}(\unitvec{s}^L),
\label{eq-app:cherepkov-expansion} 
\end{align} 
\begin{align}
A_{L,M_{L},S,M_{S}}= & \dfrac{4\pi \sqrt{2\pi}}{3}  |E_\omega^L|^{2}  \sum(-1)^{m_{2}'+\xi'-\xi+\mu_{2}'-1/2} (2J+1) \sqrt{\frac{(2\ell_1+1)(2\ell_2+1)(2L+1)}{4\pi}} \nonumber \\
&\times 
\begin{pmatrix}
	\ell_{2} & \ell_{1} & L\\
	0 & 0 & 0
\end{pmatrix}
\begin{pmatrix}
	\ell_{2} & \ell_{1} & L\\
	m_{2}' & -m_{1}' & M_{L}'
\end{pmatrix} 
\begin{pmatrix}
	1 & 1 & J\\
	\xi'' & -\xi' & -M_{J}'
\end{pmatrix}
\begin{pmatrix}
	1 & 1 & J\\
	\xi & -\xi & 0
\end{pmatrix} \nonumber \\
&\times 
\begin{pmatrix}
	\frac{1}{2} & \frac{1}{2} & S\\
	\mu_{2}' & -\mu_{1}' & M_{S}'
\end{pmatrix}
\begin{pmatrix}
	L & S & J\\
	M_{L}' & M_{S}' & M_{J}'
\end{pmatrix}
\begin{pmatrix}
	L & S & J\\
	-M_{L} & -M_{S} & 0
\end{pmatrix}
(D_{\xi'}^{\ell_1,m_1',\mu_1'})^* D_{\xi''}^{\ell_2,m_2',\mu_2'}.
\label{eq-app:cherepkovcoeff}
\end{align}
\end{subequations}
The nonzero values for $A_{L,M_{L},S,M_{S}}$ are deduced from the triangle inequality satisfied by the following 3j-symbols:
\begin{align}
\begin{pmatrix}
	1 & 1 & J\\
	\xi & -\xi & 0
\end{pmatrix} \quad , \quad 
\begin{pmatrix}
	\frac{1}{2} & \frac{1}{2} & S\\
	\mu_{2}' & -\mu_{1}' & M_{S}'
\end{pmatrix} \quad , \quad
\begin{pmatrix}
	L & S & J\\
	-M_{L} & -M_{S} & 0
\end{pmatrix}.
\end{align}
The first two 3j-symbols constrains the values of $J$, and $S$, i.e., $0\leq J \leq 2$, $0\leq S\leq 1$, respectively. This consequently constrains the value of $L$, i.e.,  $0\leq L\leq3$. Additionally, the last 3j-symbol imposes the constraint $M_L+M_S=0$. These conditions will lead to thirteen terms in the expansion Eq. \eqref{eq-app:cherepkov-expansion}, however, these terms are not independent based on the symmetry of the 3j-symbol
\begin{align}
\begin{pmatrix}
	L & S & J\\
	-M_{L} & -M_{S} & 0
\end{pmatrix}.
\end{align}
Indeed, we can show that 
\begin{align}
A_{2,0,1,0} \propto &
\begin{pmatrix}
	2 & 1 & J\\
	0 & 0 & 0
\end{pmatrix} = \sqrt{\dfrac{2}{15}}\delta_{J,1} \\
A_{2,\pm1,1,\mp1} \propto &
\begin{pmatrix}
	2 & 1 & J\\
	\pm 1 & \mp 1 & 0
\end{pmatrix} = -\dfrac{1}{\sqrt{10}}\left( \delta_{J,1} \pm \delta_{J,2} \right) \\
A_{3,0,1,0} \propto &
\begin{pmatrix}
	3 & 1 & J\\
	0 & 0 & 0
\end{pmatrix} = -\sqrt{\dfrac{3}{35}}\delta_{J,2} \\
A_{3,\pm1,1,\mp1} \propto &
\begin{pmatrix}
	3 & 1 & J\\
	\pm 1 & \mp 1 & 0
\end{pmatrix} = \sqrt{\dfrac{2}{35}}\delta_{J,2}
\end{align}
which leads to the following relations
\begin{equation}
A_{2,1,1,-1}+A_{2,-1,1,1}= -\sqrt{3}A_{2,0,1,0}
\label{eq:rel1}
\end{equation}
\begin{equation}
A_{3,1,1,-1}=A_{3,-1,1,1}=-\sqrt{\frac{2}{3}}A_{3,0,1,0}.
\label{eq:rel2}
\end{equation}

Explicit expansion of the spherical harmonics in Eq. \eqref{eq-app:cherepkov_Y} results to 
\begin{align}
W^L(\unitvec{k}^L,\unitvec{s}^L) =& \dfrac{1}{4\pi}A_{0,0,0,0} + \dfrac{\sqrt{5}}{8\pi} A_{2,0,0,0}\left(3k_z^2-1\right) + \dfrac{\sqrt{3}}{4\pi} A_{1,0,0,0} k_z + \dfrac{\sqrt{3}}{4\pi} A_{0,0,1,0} s_z \nonumber \\
&+ \dfrac{3}{4\pi}A_{1,0,1,0}k_z s_z + \dfrac{\sqrt{15}}{8\pi}A_{2,0,1,0}\left( 3k_z^2-1 \right)s_z + \dfrac{\sqrt{21}}{8\pi} A_{3,0,1,0} \left( 5k_z^2-3 \right)k_zs_z  \nonumber \\
&- \dfrac{3}{8\pi} \left( A_{1,-1,1,1} + A_{1,1,1,-1} \right) \left( k_x s_x + k_y s_y\right) - \dfrac{3i}{8\pi} \left( A_{1,-1,1,1} - A_{1,1,1,-1} \right) \left( s_y k_x - s_x k_y \right) \nonumber \\
&-\dfrac{3\sqrt{5}}{8\pi}  \left( A_{2,-1,1,1} + A_{2,1,1,-1} \right) \left( k_x s_x + k_y s_y\right) k_z - \dfrac{3i\sqrt{5}}{8\pi}  \left( A_{2,-1,1,1} - A_{2,1,1,-1} \right) \left( s_y k_x - s_x k_y \right) k_z \nonumber \\
&-\dfrac{3}{16\pi}\sqrt{\dfrac{7}{2}}  \left( A_{3,-1,1,1} + A_{3,1,1,-1} \right) \left( k_x s_x + k_y s_y\right) \left( 5k_z^2-1 \right)\nonumber \\
&-\dfrac{3i}{16\pi}\sqrt{\dfrac{7}{2}}  \left( A_{3,-1,1,1} - A_{3,1,1,-1} \right) \left( s_y k_x - s_x k_y \right) \left( 5k_z^2-1 \right)
\label{eq-app:expansion}
\end{align}
Using the relations Eqs. \eqref{eq:rel1}-\eqref{eq:rel2}, we can regroup the following terms:
\begin{subequations}
\begin{align}
\dfrac{\sqrt{15}}{8\pi}&A_{2,0,1,0}\left( 3k_z^2-1 \right)s_z -\dfrac{3\sqrt{5}}{8\pi}  \left( A_{2,-1,1,1} + A_{2,1,1,-1} \right) \left( k_x s_x + k_y s_y\right) k_z \nonumber \\
=& \dfrac{\sqrt{15}}{8\pi} A_{2,0,1,0} \left[ \left( 3k_z^2-1 \right)s_z + 3\left( k_x s_x + k_y s_y\right) k_z \right] 
= \dfrac{\sqrt{15}}{8\pi} A_{2,0,1,0} \left[ 3\left( k_x s_x + k_y s_y + k_zs_z\right) k_z - s_z \right]
\end{align}
\begin{align}
\dfrac{\sqrt{21}}{8\pi} & A_{3,0,1,0} \left( 5k_z^2-3 \right)k_zs_z  -\dfrac{3}{16\pi}\sqrt{\dfrac{7}{2}}  \left( A_{3,-1,1,1} + A_{3,1,1,-1} \right) \left( k_x s_x + k_y s_y\right) \left( 5k_z^2-1 \right) \nonumber \\
=& \dfrac{\sqrt{21}}{8\pi} A_{3,0,1,0} \left[ \left( 5k_z^2-3 \right)k_zs_z + \left( k_x s_x + k_y s_y\right) \left( 5k_z^2-1 \right)\right] =  \dfrac{\sqrt{21}}{8\pi} A_{3,0,1,0} \left[ \left( 5k_z^2-1 \right)\left( k_x s_x + k_y s_y + k_z s_z \right) - 2k_zs_z  \right]
\end{align}
\end{subequations}
Finally, we rearrange Eq. \eqref{eq-app:expansion} to obtain the vector form: 
\begin{align}
W^{L}(\unitvec{k}^{L},\unitvec{s}^{L}) 
=& \left[ \dfrac{1}{8\pi}\left(2A_{0,0,0,0}\right) \right] \bigg\{  1 - \dfrac{1}{2}\left(-\sqrt{5}\dfrac{A_{2,0,0,0}}{A_{0,0,0,0}}\right) \left[ 3(\unitvec{k}\cdot\unitvec{z})^2-1 \right] + \left( \sqrt{3} \dfrac{A_{0,0,1,0}}{A_{0,0,0,0}} \right) (\unitvec{s}\cdot\unitvec{z}) \bigg. \nonumber \\
&- \left[ \dfrac{3i\sqrt{5}}{2} \left( \dfrac{A_{2,-1,1,1}-A_{2,1,1,-1}}{A_{0,0,0,0}} \right) \right] (\unitvec{z} \cdot \unitvec{s} \times \unitvec{k}) (\unitvec{k} \cdot \unitvec{z}) 
-\left( -\sqrt{15}\dfrac{A_{2,0,1,0}}{A_{0,0,0,0}} \right) \left[ \dfrac{3}{2} (\unitvec{k}\cdot\unitvec{s})(\unitvec{k}\cdot\unitvec{z}) - \dfrac{1}{2}(\unitvec{s} \cdot \unitvec{z}) \right] \nonumber \\
&+ \left( \sqrt{3} \dfrac{A_{1,0,0,0}}{A_{0,0,0,0}} \right) (\unitvec{k}\cdot\unitvec{z}) + \left[ - \dfrac{3i}{2} \left( \dfrac{A_{1,-1,1,1}-A_{1,1,1,-1}}{A_{0,0,0,0}} \right) \right] ( \unitvec{z} \cdot \unitvec{s} \times \unitvec{k} ) \nonumber \\
&+ \left[ -\dfrac{3}{2} \left( \dfrac{ (A_{1,-1,1,} + A_{1,1,1,-1} ) + \sqrt{\frac{7}{3}} A_{3,0,1,0} }{ A_{0,0,0,0} } \right) \right] (\unitvec{k}\cdot\unitvec{s}) \nonumber \\
&+ \big. \left[ \dfrac{3A_{1,0,1,0}+\frac{3}{2}(A_{1,-1,1,1}+A_{1,1,1,-1})-\sqrt{21}A_{3,0,1,0}}{A_{0,0,0,0}} \right] (\unitvec{k}\cdot\unitvec{z})(\unitvec{s}\cdot\unitvec{z}) +\left( \dfrac{5\sqrt{21}}{2} \dfrac{A_{3,0,1,0}}{A_{0,0,0,0}} \right)   (\unitvec{k}\cdot\unitvec{z})^2 (\unitvec{k}\cdot\unitvec{s}) \bigg\}
\label{eq-app:cherepkov-vectors}
\end{align}
Finally, we compare with the original expression of Cherepkov \cite{cherepkov1983manifestations}:

\begin{subequations} \label{eq-app:parameters}
\begin{minipage}{0.40\textwidth}
\begin{align}
\sigma_{\text{cross}} =& 2 \, A_{0,0,0,0} 
\end{align}
\begin{align}
\beta =& -\sqrt{5} \, \frac{A_{2,0,0,0} }{A_{0,0,0,0}} 
\end{align}
\begin{align}
A =& \sqrt{3} \, \frac{A_{0,0,1,0}}{A_{0,0,0,0}} 
\end{align}
\begin{align}
\eta=&i\frac{3\sqrt{5}}{2} \, \frac{(A_{2,-1,1,1}-A_{2,1,1,-1})}{A_{0,0,0,0}} 
\end{align}
\begin{align}
\gamma =&-\sqrt{15} \, \frac{A_{2,0,1,0}}{A_{0,0,0,0}} 
\end{align}
\end{minipage}
\hfill
\begin{minipage}{0.55\textwidth}
\begin{align}
D=&\sqrt{3} \, \frac{A_{1,0,0,0}}{A_{0,0,0,0}} 
\end{align}
\begin{align}
C =& -i \frac{3}{2} \, \frac{(A_{1,-1,1,1}-A_{1,1,1,-1})}{A_{0,0,0,0}} 
\end{align}
\begin{align}
B_1 =& -\dfrac{3}{2} \left[\dfrac{(A_{1,1,1,-1}+A_{1,-1,1,1})+\sqrt{\frac{7}{3}}A_{3,0,1,0}}{A_{0,0,0,0}}\right] 
\end{align}
\begin{align}
B_2 =& \dfrac{3A_{1,0,1,0} + \frac{3}{2}(A_{1,1,1,-1}+A_{1,-1,1,1}) - \sqrt{21}A_{3,0,1,0}}{A_{0,0,0,0}} 
\end{align}
\begin{align}
B_3 =& \frac{5\sqrt{21}}{2} \, \frac{A_{3,0,1,0}}{A_{0,0,0,0}} 
\end{align}
\end{minipage}
\end{subequations}

Using Eq. \eqref{eq-app:cherepkov-vectors}, we find that there are errors with the the parameters $\{\eta,D,C,B_1,B_2\}$ listed in Ref \cite{cherepkov1983manifestations}:
\begin{minipage}{0.4\textwidth}
\begin{align}
\tilde{\eta} = -i\frac{3\sqrt{5}}{2} \dfrac{(A_{2,-1,1,1}-A_{2,1,1,-1}}{A_{0,0,0,0}}
\end{align}
\begin{align}
\tilde{D} = \dfrac{A_{1,0,0,0}}{A_{0,0,0,0}}
\end{align}
\begin{align}
\tilde{C} = i \frac{3}{2} \, \frac{(A_{1,-1,1,1}-A_{1,1,1,-1})}{A_{0,0,0,0}} 
\end{align}
\end{minipage}
\begin{minipage}{0.55\textwidth}
\begin{align}
\tilde{B}_1 =  -\dfrac{3}{2} \left[\dfrac{(A_{1,1,1,-1}+A_{1,-1,1,1})-\sqrt{\frac{7}{3}}A_{3,0,1,0}}{A_{0,0,0,0}}\right] 
\end{align}
\begin{align}
\tilde{B}_2 =& \dfrac{A_{1,0,1,0} + \frac{3}{2}(A_{1,1,1,-1}+A_{1,-1,1,1}) - \sqrt{21}A_{3,0,1,0}}{A_{0,0,0,0}} 
\end{align}
\end{minipage}

The rearrangement of terms is completely arbitrary, however, as was shown in Sec. \ref{sec:vector-form}, the coefficients $A_{L,M_{L},S,M_{S}}$
are examples of tensors of rank $L+S$, which in principle, can be directly measured using a detector setup with a structure that reflects the corresponding $Y_{L,M_L}Y_{S,M_S}$. Now, the parameters Eqs. \eqref{eq:cherepkov-sigma}-\eqref{eq:cherepkov-B3} used by Cherepkov are formed from combinations of the expansion coefficients $A_{L,M_{L},S,M_{S}}$ which means that measuring some of these parameters will therefore require several detector setups to measure each $A_{L,M_{L},S,M_{S}}$.


\section{Full details on the vector form of the photoionization parameters}
\label{app:vectors}

We provide the full details of the calculation leading to Eq. \eqref{eq:dynamic} of the main text. Using Eq. \eqref{eq:dme}, the photoionization yield is now  
\begin{align}
W^{L} (\unitvec{k}^{L},\unitvec{s}^{L}) = & \int d\rho\sum_{\mu_{1}^L,\mu_{2}^L}\sum_{\mu_{1}^M,\mu_{2}^M}(\fullvec{D}_{\fullvec{k}^M\mu_{1}^M}^{L}\cdot\fullvec{E}^{L})\mathscr{D}_{\mu_{1}^M,\mu_{1}^L}^{1/2*}\left(\dfrac{\mathbb{I}+\unitvec{s}^{L}\cdot\unitvec{\sigma}^L}{2}\right)_{\mu_{1}^L,\mu_{2}^L}\mathscr{D}_{\mu_{2}^M,\mu_{2}^L}^{1/2}(\fullvec{D}_{\fullvec{k}^M,\mu_{2}^M}^{L*}\cdot\fullvec{E}^{L*}).
\label{eq:app_yieldfull}
\end{align}
The relevant quantity to calculate is the product: 
\begin{align}
\sum_{\mu_{1}^L,\mu_{2}^L} \mathscr{D}_{\mu_{1}^M,\mu_{1}^L}^{1/2*} &  \left(\dfrac{\mathbb{I}+\unitvec{s}^{L}\cdot\unitvec{\sigma}^L}{2}\right)_{\mu_{1}^L,\mu_{2}^L}\mathscr{D}_{\mu_{2}^M,\mu_{2}^L}^{1/2} \nonumber \\
=& \sqrt{\pi} 
\begin{bmatrix}
1 & 0 \\
0 & 1
\end{bmatrix}_{\mu_1^M,\mu_2^M} Y_{0,0}(\unitvec{s}^L) + \sqrt{\dfrac{\pi}{3}}
\begin{bmatrix}
\cos\beta & -e^{i\gamma}\sin\beta \\
-e^{-i\gamma}\sin\beta & -\cos\beta
\end{bmatrix}_{\mu_1^M,\mu_2^M} Y_{1,0}(\unitvec{s}^L) \nonumber \\
&- \sqrt{\dfrac{\pi}{6}}
\begin{bmatrix}
e^{-i\alpha}\sin\beta               &       e^{-i(\alpha-\gamma)}(-1+\cos\beta) \\
e^{-i(\alpha+\gamma)}(1+\cos\beta)  &       -e^{-i\alpha}\sin\beta
\end{bmatrix}_{\mu_1^M,\mu_2^M} Y_{1,1}(\unitvec{s}^L) \nonumber \\
&+ \sqrt{\dfrac{\pi}{6}}
\begin{bmatrix}
e^{i\alpha}\sin\beta                &       e^{i(\alpha+\gamma)}(1+\cos\beta) \\
e^{i(\alpha-\gamma)}(-1+\cos\beta)  &       -e^{i\alpha}\sin\beta
\end{bmatrix}_{\mu_1^M,\mu_2^M} Y_{1,-1}(\unitvec{s}^L),
\label{eq:app_matrix}
\end{align}
which follows from Eqs. \eqref{eq:wigner-1/2} and \eqref{eq:projectorYs}. The expression Eq. \eqref{eq:app_matrix} can then be simplified by rewriting the matrix elements as scalar products of a molecular vector $\unitvec{a}^M$ and an effective electric field $\unitvec{\epsilon}^L$, i.e., $(\unitvec{a}^L\cdot\unitvec{\epsilon}^L)=[(R_\rho \unitvec{a}^M)\cdot\unitvec{\epsilon}^L]$, where, $R_\rho$ is the Euler rotation matrix (zyz-convention):
\begin{equation}
R_{\rho}=\begin{bmatrix}-\sin\alpha\sin\gamma+\cos\alpha\cos\beta\cos\gamma & -\sin\alpha\cos\gamma-\cos\alpha\cos\beta\sin\gamma & \cos\alpha\sin\beta\\
\cos\alpha\sin\gamma+\sin\alpha\cos\beta\cos\gamma & \cos\alpha\cos\gamma-\sin\alpha\cos\beta\sin\gamma & \sin\alpha\sin\beta\\
-\sin\beta\cos\gamma & \sin\beta\sin\gamma & \cos\beta
\end{bmatrix},
\end{equation}
that rotates vectors from the molecular frame to the laboratory frame. We denote the molecular $x,y,z$-axes by the unit vectors $\{ \unitvec{\chi}^M , \unitvec{\eta}^M , \unitvec{\zeta}^M \}$, and the effective electric fields as
\begin{equation}
\unitvec{\epsilon}_{0}^L = \unitvec{z}^L \quad , \quad \unitvec{\epsilon}_{\pm}^L = \dfrac{\mp (\unitvec{x}^L \pm i \unitvec{y}^L)}{\sqrt{2}},
\end{equation}       
where, the unit vectors $\{ \unitvec{x}^L , \unitvec{y}^L , \unitvec{z}^L \}$ denote the laboratory $x,y,z$-axes. It easily follows that 
\begin{equation}
\cos\beta = (R_\rho \unitvec{\zeta}^M) \cdot \unitvec{\epsilon}_0^L = \unitvec{\zeta}^L \cdot \unitvec{\epsilon}_0^L
\end{equation}
\begin{equation}
e^{\pm i\gamma} \sin\beta = - \left[ R_\rho \left( \unitvec{\chi}^M \mp i \unitvec{\eta}^M \right) \right] \cdot  \unitvec{\epsilon}_0^L = - \left( \unitvec{\chi}^L \mp i \unitvec{\eta}^L \right) \cdot  \unitvec{\epsilon}_0^L
\end{equation}
\begin{equation}
e^{\pm i \alpha} \sin\beta = \mp \sqrt{2} \left( R_\rho \unitvec{\zeta}^M \right) \cdot \unitvec{\epsilon}_\pm^L = \mp \sqrt{2} \left(   \unitvec{\zeta}^L \cdot \unitvec{\epsilon}_\pm^L \right)
\end{equation}
\begin{equation}
e^{\pm i (\alpha + \gamma)} (1+\cos\beta) = \mp \sqrt{2} \left[ R_\rho \left( \unitvec{\chi}^M \mp i \unitvec{\eta}^M \right) \right] \cdot \unitvec{\epsilon}_\pm^L = \mp \sqrt{2} \left( \unitvec{\chi}^L \mp i \unitvec{\eta}^L \right) \cdot \unitvec{\epsilon}_\pm^L
\end{equation}
\begin{equation}
e^{\pm i (\alpha - \gamma)} (-1+\cos\beta) = \mp \sqrt{2} \left[ R_\rho \left( \unitvec{\chi}^M \pm i \unitvec{\eta}^M \right) \right] \cdot \unitvec{\epsilon}_\pm^L = \mp \sqrt{2} \left( \unitvec{\chi}^L \pm i \unitvec{\eta}^L \right) \cdot \unitvec{\epsilon}_\pm^L
\end{equation}
such that we can now express Eq. \eqref{eq:app_matrix} as 
\begin{align}
\sum_{\mu_{1}^L,\mu_{2}^L} \mathscr{D}_{\mu_{1}^M,\mu_{1}^L}^{1/2*} &  \left(\dfrac{\mathbb{I}+\unitvec{s}^{L}\cdot\unitvec{\sigma}^L}{2}\right)_{\mu_{1}^L,\mu_{2}^L}\mathscr{D}_{\mu_{2}^M,\mu_{2}^L}^{1/2} \nonumber \\
=& \sqrt{\pi} 
\begin{bmatrix}
1 & 0 \\
0 & 1
\end{bmatrix}_{\mu_1^M,\mu_2^M} Y_{0,0}(\unitvec{s}^L) + \sqrt{\dfrac{\pi}{3}}
\left\{ 
\begin{bmatrix}
    \unitvec{\zeta}^L                       &       \unitvec{\chi}^L - i \unitvec{\eta}^L \\
    \unitvec{\chi}^L + i \unitvec{\eta}^L   &       - \unitvec{\zeta}^L 
\end{bmatrix}
\cdot \unitvec{\epsilon}_0^L\right\}_{\mu_1^M,\mu_2^M} Y_{1,0}(\unitvec{s}^L) \nonumber \\
&- \sqrt{\dfrac{\pi}{3}}
\left\{ 
\begin{bmatrix}
    \unitvec{\zeta}^L                       &       \unitvec{\chi}^L - i \unitvec{\eta}^L \\
    \unitvec{\chi}^L + i \unitvec{\eta}^L   &       - \unitvec{\zeta}^L 
\end{bmatrix}
\cdot \unitvec{\epsilon}_{-}^L\right\}_{\mu_1^M,\mu_2^M} Y_{1,1}(\unitvec{s}^L)  \nonumber \\
&- \sqrt{\dfrac{\pi}{3}}
\left\{ 
\begin{bmatrix}
    \unitvec{\zeta}^L                       &       \unitvec{\chi}^L - i \unitvec{\eta}^L \\
    \unitvec{\chi}^L + i \unitvec{\eta}^L   &       - \unitvec{\zeta}^L 
\end{bmatrix}
\cdot \unitvec{\epsilon}_{+}^L\right\}_{\mu_1^M,\mu_2^M} Y_{1,-1}(\unitvec{s}^L) \nonumber \\
=& \sqrt{\pi} \delta_{\mu_1^M,\mu_2^M}  Y_{0,0}(\unitvec{s}^L) + \sqrt{\dfrac{\pi}{3}} \left( \unitvec{\sigma}^L_{\mu_1^M,\mu_2^M} \cdot \unitvec{\epsilon}_0^L \right)  Y_{1,0}(\unitvec{s}^L) \nonumber \\
&- \sqrt{\dfrac{\pi}{3}} \left( \unitvec{\sigma}^L_{\mu_1^M,\mu_2^M} \cdot \unitvec{\epsilon}_{-}^L \right)  Y_{1,1}(\unitvec{s}^L) - \sqrt{\dfrac{\pi}{3}} \left( \unitvec{\sigma}^L_{\mu_1^M,\mu_2^M} \cdot \unitvec{\epsilon}_{+}^L \right)  Y_{1,-1}(\unitvec{s}^L)
\label{eq:app_matrix_2}
\end{align}
where, we introduced the vector of Pauli spin matrices $\unitvec{\sigma}^L=\sigma_x\unitvec{\chi}^L+\sigma_y\unitvec{\eta}^L+\sigma_z\unitvec{\zeta}^L$. Substituting Eq. \eqref{eq:app_matrix_2} into Eq. \eqref{eq:app_yieldfull}, we obtain a compact vector expression of the yield: 
\begin{subequations}
\begin{equation}
W^L(\unitvec{k}^L,\unitvec{s}^L,\rho) = \sum_{\ell_s,m_s}  \mathcal{W}_{\ell_s,m_s}^L(\unitvec{k}^L,\rho)Y_{\ell_s,m_s}(\unitvec{s}^L)
\end{equation}
\begin{equation}
\mathcal{W}_{\ell_s,m_s}^L(\unitvec{k}^L,\rho) = \sum_{\mu_1^M,\mu_2^M} (-1)^{m_s} \sqrt{\dfrac{\pi}{2\ell_s+1}} \, \Lambda^{(\ell_s,m_s)}_{\mu_1^M,\mu_2^M} \left( \fullvec{D}_{\fullvec{k}^M,\mu_1^M}^{L}\cdot\fullvec{E}^{L} \right) \left( \fullvec{D}_{\fullvec{k}^M,\mu_2^M}^{L*}\cdot\fullvec{E}^{L*} \right) 
\end{equation}
\begin{align}
\Lambda^{(\ell_s,m_s)}_{\mu_1^M,\mu_2^M} = 
\begin{cases}
    \delta_{\mu_1^M,\mu_2^M} \quad ; & \ell_s = 0 \\
    \left( \unitvec{\sigma}_{\mu_1^M,\mu_2^M}^L\cdot\unitvec{\epsilon}_{-m_s}^L \right)  \quad ;& \ell_s =1 
\end{cases}
\end{align}
\end{subequations}

Since $W^{L}(\unitvec{k}^{L},\unitvec{s}^{L})$ depends on two directions, we can equivalently perform an expansion 
\begin{align}
W^L(\unitvec{k}^L,\unitvec{s}^L) = \sum b_{\ell,m_{\ell},\ell_{s},m_{s}}Y_{\ell,m_{\ell}}(\unitvec{k}^{L})Y_{\ell_{s},m_{s}}(\unitvec{s}^{L})
\label{eq:app_yield}
\end{align}
wherein, the coefficients $b_{l,m_{l},l_{s},m_{s}}$ now encode any information about the molecule and ionizing field, and have the form 
\begin{align}
b_{l,m_{l},l_{s},m_{s}} 
=&  (-1)^{m_s}\sqrt{ \dfrac{\pi}{2\ell_s+1} } \sum_{\mu_1^M,\mu_2^M} \int d\Theta_k^M   \int d\rho  \left( \fullvec{D}_{\fullvec{k}^M,\mu_1^M}^{L*}\cdot\fullvec{E}^{L*} \right) \left( \fullvec{D}_{\fullvec{k}^M,\mu_2^M}^{L}\cdot\fullvec{E}^{L} \right) \Lambda^{(l_s,m_s)}_{\mu_2^M,\mu_1^M}  Y_{l,m_{l}}^*(\unitvec{k}^{L})
\label{eq:app_coeff_b}
\end{align}
It follows from Ref. \cite{andrews1977three} that performing the orientation averaging in Eq. \eqref{eq:app_coeff_b} will yield the general form $\sum_{ij}g_i M_{ij}f_j$, where, $g_i$ and $f_j$ are rotational invariants that are formed from the set of vectors fixed in the molecular and laboratory frame, respectively, while $M_{ij}$ are coupling constants between the molecular and rotational invariants. The vectors in Eq. \eqref{eq:app_coeff_b} that are fixed in the molecular frame are $\{\fullvec{D}_{\fullvec{k}^M,\mu^M}^{M} , \unitvec{\sigma}_{\mu^M,\mu^M}^M, \unitvec{k}^M\}$ which appear above to be rotated into the laboratory frame ($\fullvec{a}^L=R_\rho\fullvec{a}^M$), while the rest are fixed in the laboratory frame. Note that the transition dipoles $\fullvec{D}_{\fullvec{k}^M,\mu^M}^M$ is a molecular property and therefore rotates with the molecular frame while $\mu^M$ refers to either spin up or down in z-axis of the molecular frame. Thus, to rotate $\fullvec{D}_{\fullvec{k}^M,\mu^M}^M$ from the molecular frame to the laboratory frame we have $\fullvec{D}_{\fullvec{k}^M,\mu^M}^L=R_\rho\fullvec{D}_{\fullvec{k}^M,\mu^M}^M$. Similarly, we have $\unitvec{\sigma}_{\mu^M,\mu^M}^L=R_\rho\unitvec{\sigma}_{\mu^M,\mu^M}^M$.

To evaluate the expansion coefficients $b_{l,m_l,l_s,m_s}$'s, we use the following vector identities from Ref. \cite{andrews1977three}:    
\begin{align}
\int d\rho (\fullvec{a}^L\cdot\fullvec{r}^L)(\fullvec{b}^L\cdot\fullvec{s}^L)=& \dfrac{1}{3} \left( \fullvec{a}^M \cdot \fullvec{b}^M\right) \left( \fullvec{r}^L \cdot \fullvec{s}^L \right).
\label{eq:rank2}
\end{align} 
\begin{align}
\int d\rho (\fullvec{a}^L\cdot\fullvec{r}^L)(\fullvec{b}^L\cdot\fullvec{s}^L)(\fullvec{c}^L\cdot\fullvec{t}^L) =& \dfrac{1}{6} \left( \fullvec{a}^M \cdot \fullvec{b}^M \times \fullvec{c}^M \right) \left( \fullvec{r}^L \cdot \fullvec{s}^L \times \fullvec{t}^L \right).
\label{eq:rank3}
\end{align} 
\begin{align}
\int d\rho (\fullvec{a}^L\cdot\fullvec{r}^L)(\fullvec{b}^L\cdot\fullvec{s}^L)(\fullvec{c}^L\cdot\fullvec{t}^L) (\fullvec{d}^L\cdot\fullvec{u}^L) 
%
%
%
=& \dfrac{1}{30} 
\begin{bmatrix}
(\fullvec{a}^M\cdot\fullvec{b}^M)(\fullvec{c}^M\cdot\fullvec{d}^M) \\
(\fullvec{a}^M\cdot\fullvec{c}^M)(\fullvec{b}^M\cdot\fullvec{d}^M) \\
(\fullvec{a}^M\cdot\fullvec{d}^M)(\fullvec{b}^M\cdot\fullvec{c}^M)
\end{bmatrix}^T
\begin{bmatrix}
4 & -1 & -1 \\
-1 & 4 & -1 \\
-1 & -1 & 4 \\
\end{bmatrix}
\begin{bmatrix}
(\fullvec{r}^L \cdot \fullvec{s}^L)(\fullvec{t}^L \cdot \fullvec{u}^L) \\
(\fullvec{r}^L \cdot \fullvec{t}^L)(\fullvec{s}^L \cdot \fullvec{u}^L) \\
(\fullvec{r}^L \cdot \fullvec{u}^L)(\fullvec{s}^L \cdot \fullvec{u}^L) 
\end{bmatrix}.
\label{eq:rank4}
\end{align}
\begin{align}
\int d\rho & (\fullvec{a}^L\cdot\fullvec{r}^L)(\fullvec{b}^L\cdot\fullvec{s}^L)(\fullvec{c}^L\cdot\fullvec{t}^L) (\fullvec{d}^L\cdot\fullvec{u}^L) (\fullvec{f}^L\cdot\fullvec{v}^L) \nonumber \\
=& \dfrac{1}{30} \left\{ (\fullvec{a}^M \cdot \fullvec{b}^M \times \fullvec{c}^M)(\fullvec{d}^M\cdot \fullvec{f}^M)(\fullvec{r}^L \cdot \fullvec{s}^L \times \fullvec{t}^L)(\fullvec{u}^L\cdot\fullvec{v}^L) + (\fullvec{a}^M \cdot \fullvec{b}^M \times \fullvec{d}^M)(\fullvec{c}^M\cdot \fullvec{f}^M)(\fullvec{r}^L \cdot \fullvec{s}^L \times \fullvec{u}^L)(\fullvec{t}^L\cdot\fullvec{v}^L)   \right. \nonumber \\
& + (\fullvec{a}^M \cdot \fullvec{b}^M \times \fullvec{f}^M)(\fullvec{c}^M\cdot \fullvec{d}^M)(\fullvec{r}^L \cdot \fullvec{s}^L \times \fullvec{v}^L)(\fullvec{t}^L\cdot\fullvec{u}^L) + (\fullvec{a}^M \cdot \fullvec{c}^M \times \fullvec{d}^M)(\fullvec{b}^M\cdot \fullvec{f}^M)(\fullvec{r}^L \cdot \fullvec{t}^L \times \fullvec{u}^L)(\fullvec{s}^L\cdot\fullvec{v}^L) \nonumber \\
& + (\fullvec{a}^M \cdot \fullvec{c}^M \times \fullvec{f}^M)(\fullvec{b}^M\cdot \fullvec{d}^M)(\fullvec{r}^L \cdot \fullvec{t}^L \times \fullvec{v}^L)(\fullvec{s}^L\cdot\fullvec{u}^L) + (\fullvec{a}^M \cdot \fullvec{d}^M \times \fullvec{f}^M)(\fullvec{b}^M\cdot \fullvec{c}^M)(\fullvec{r}^L \cdot \fullvec{u}^L \times \fullvec{v}^L)(\fullvec{s}^L\cdot\fullvec{t}^L) \nonumber \\
& + (\fullvec{b}^M \cdot \fullvec{c}^M \times \fullvec{d}^M)(\fullvec{a}^M\cdot \fullvec{f}^M)(\fullvec{s}^L \cdot \fullvec{t}^L \times \fullvec{u}^L)(\fullvec{r}^L\cdot\fullvec{v}^L)  + (\fullvec{b}^M \cdot \fullvec{c}^M \times \fullvec{f}^M)(\fullvec{a}^M\cdot \fullvec{d}^M)(\fullvec{s}^L \cdot \fullvec{t}^L \times \fullvec{v}^L)(\fullvec{r}^L\cdot\fullvec{u}^L) \nonumber \\
& + \left.  (\fullvec{b}^M \cdot \fullvec{d}^M \times \fullvec{f}^M)(\fullvec{a}^M\cdot \fullvec{c}^M)(\fullvec{s}^L \cdot \fullvec{u}^L \times \fullvec{v}^L)(\fullvec{r}^L\cdot\fullvec{t}^L) + (\fullvec{c}^M \cdot \fullvec{d}^M \times \fullvec{f}^M)(\fullvec{a}^M\cdot \fullvec{b}^M)(\fullvec{t}^L \cdot \fullvec{u}^L \times \fullvec{v}^L)(\fullvec{r}^L\cdot\fullvec{s}^L)\right\}.
\label{eq:rank5}
\end{align}
\begin{subequations}
\begin{align}
\int d\rho (\fullvec{a}^L\cdot\fullvec{r}^L)(\fullvec{b}^L\cdot\fullvec{s}^L)(\fullvec{c}^L\cdot\fullvec{t}^L) (\fullvec{d}^L\cdot\fullvec{u}^L) (\fullvec{f}^L\cdot\fullvec{v}^L) (\fullvec{g}^L\cdot\fullvec{w}^L) =  \fullvec{G}^{(6)}\cdot\mathcal{M}^{(6)}\fullvec{F}^{(6)}
\label{eq:rank6}
\end{align}
\begin{align}
\mathcal{M}^{(6)} = \dfrac{1}{210}
\begin{bmatrix}
16 & -5 & -5 & -5 & 2 & 2 & -5 & 2 & 2 & 2 & 2 & -5 & 2 & 2 & -5\\
-5 & 16 & -5 & 2 & -5 & 2 & 2 & 2 & -5 & -5 & 2 & 2 & 2 & -5 & 2\\
-5 & -5 & 16 & 2 & 2 & -5 & 2 & -5 & 2 & 2 & -5 & 2 & -5 & 2 & 2\\
-5 & 2 & 2 & 16 & -5 & -5 & -5 & 2 & 2 & 2 & -5 & 2 & 2 & -5 & 2\\
2 & -5 & 2 & -5 & 16 & -5 & 2 & -5 & 2 & -5 & 2 & 2 & 2 & 2 & -5\\
2 & 2 & -5 & -5 & -5 & 16 & 2 & 2 & -5 & 2 & 2 & -5 & -5 & 2 & 2\\
-5 & 2 & 2 & -5 & 2 & 2 & 16 & -5 & -5 & -5 & 2 & 2 & -5 & 2 & 2\\
2 & 2 & -5 & 2 & -5 & 2 & -5 & 16 & -5 & 2 & -5 & 2 & 2 & 2 & -5\\
2 & -5 & 2 & 2 & 2 & -5 & -5 & -5 & 16 & 2 & 2 & -5 & 2 & -5 & 2\\
2 & -5 & 2 & 2 & -5 & 2 & -5 & 2 & 2 & 16 & -5 & -5 & -5 & 2 & 2\\
2 & 2 & -5 & -5 & 2 & 2 & 2 & -5 & 2 & -5 & 16 & -5 & 2 & -5 & 2\\
-5 & 2 & 2 & 2 & 2 & -5 & 2 & 2 & -5 & -5 & -5 & 16 & 2 & 2 & -5\\
2 & 2 & -5 & 2 & 2 & -5 & -5 & 2 & 2 & -5 & 2 & 2 & 16 & -5 & -5\\
2 & -5 & 2 & -5 & 2 & 2 & 2 & 2 & -5 & 2 & -5 & 2 & -5 & 16 & -5\\
-5 & 2 & 2 & 2 & -5 & 2 & 2 & -5 & 2 & 2 & 2 & -5 & -5 & -5 & 16
\end{bmatrix}
\label{eq:rank6M}
\end{align}
\begin{align}
\fullvec{G}^{(6)} =
\begin{bmatrix}
(\fullvec{a}^M\cdot\fullvec{b}^M)(\fullvec{c}^M\cdot\fullvec{d}^M) (\fullvec{f}^M\cdot\fullvec{g}^M) \\
(\fullvec{a}^M\cdot\fullvec{b}^M)(\fullvec{c}^M\cdot\fullvec{f}^M) (\fullvec{d}^M\cdot\fullvec{g}^M) \\
(\fullvec{a}^M\cdot\fullvec{b}^M)(\fullvec{c}^M\cdot\fullvec{g}^M) (\fullvec{d}^M\cdot\fullvec{f}^M) \\
(\fullvec{a}^M\cdot\fullvec{c}^M)(\fullvec{b}^M\cdot\fullvec{d}^M) (\fullvec{f}^M\cdot\fullvec{g}^M) \\
(\fullvec{a}^M\cdot\fullvec{c}^M)(\fullvec{b}^M\cdot\fullvec{f}^M) (\fullvec{d}^M\cdot\fullvec{g}^M) \\
(\fullvec{a}^M\cdot\fullvec{c}^M)(\fullvec{b}^M\cdot\fullvec{g}^M) (\fullvec{d}^M\cdot\fullvec{f}^M) \\
(\fullvec{a}^M\cdot\fullvec{d}^M)(\fullvec{b}^M\cdot\fullvec{c}^M) (\fullvec{f}^M\cdot\fullvec{g}^M) \\
(\fullvec{a}^M\cdot\fullvec{d}^M)(\fullvec{b}^M\cdot\fullvec{f}^M) (\fullvec{c}^M\cdot\fullvec{g}^M) \\
(\fullvec{a}^M\cdot\fullvec{d}^M)(\fullvec{b}^M\cdot\fullvec{g}^M) (\fullvec{c}^M\cdot\fullvec{f}^M) \\
(\fullvec{a}^M\cdot\fullvec{f}^M)(\fullvec{b}^M\cdot\fullvec{c}^M) (\fullvec{d}^M\cdot\fullvec{g}^M) \\
(\fullvec{a}^M\cdot\fullvec{f}^M)(\fullvec{b}^M\cdot\fullvec{d}^M) (\fullvec{c}^M\cdot\fullvec{g}^M) \\
(\fullvec{a}^M\cdot\fullvec{f}^M)(\fullvec{b}^M\cdot\fullvec{g}^M) (\fullvec{c}^M\cdot\fullvec{d}^M) \\
(\fullvec{a}^M\cdot\fullvec{g}^M)(\fullvec{b}^M\cdot\fullvec{c}^M) (\fullvec{d}^M\cdot\fullvec{f}^M) \\
(\fullvec{a}^M\cdot\fullvec{g}^M)(\fullvec{b}^M\cdot\fullvec{d}^M) (\fullvec{c}^M\cdot\fullvec{f}^M) \\
(\fullvec{a}^M\cdot\fullvec{g}^M)(\fullvec{b}^M\cdot\fullvec{f}^M) (\fullvec{c}^M\cdot\fullvec{d}^M) 
\end{bmatrix}^T \qquad 
\fullvec{F}^{(6)} =
\begin{bmatrix}
(\fullvec{r}^L\cdot\fullvec{s}^L)(\fullvec{t}^L\cdot\fullvec{u}^L)(\fullvec{v}^L\cdot\fullvec{w}^L) \\
(\fullvec{r}^L\cdot\fullvec{s}^L)(\fullvec{t}^L\cdot\fullvec{v}^L)(\fullvec{u}^L\cdot\fullvec{w}^L) \\
(\fullvec{r}^L\cdot\fullvec{s}^L)(\fullvec{t}^L\cdot\fullvec{w}^L)(\fullvec{u}^L\cdot\fullvec{v}^L) \\
(\fullvec{r}^L\cdot\fullvec{t}^L)(\fullvec{s}^L\cdot\fullvec{u}^L)(\fullvec{v}^L\cdot\fullvec{w}^L) \\
(\fullvec{r}^L\cdot\fullvec{t}^L)(\fullvec{s}^L\cdot\fullvec{v}^L)(\fullvec{u}^L\cdot\fullvec{w}^L) \\
(\fullvec{r}^L\cdot\fullvec{t}^L)(\fullvec{s}^L\cdot\fullvec{w}^L)(\fullvec{u}^L\cdot\fullvec{v}^L) \\
(\fullvec{r}^L\cdot\fullvec{u}^L)(\fullvec{s}^L\cdot\fullvec{t}^L)(\fullvec{v}^L\cdot\fullvec{w}^L) \\
(\fullvec{r}^L\cdot\fullvec{u}^L)(\fullvec{s}^L\cdot\fullvec{v}^L)(\fullvec{t}^L\cdot\fullvec{w}^L) \\
(\fullvec{r}^L\cdot\fullvec{u}^L)(\fullvec{s}^L\cdot\fullvec{w}^L)(\fullvec{t}^L\cdot\fullvec{v}^L) \\
(\fullvec{r}^L\cdot\fullvec{v}^L)(\fullvec{s}^L\cdot\fullvec{t}^L)(\fullvec{u}^L\cdot\fullvec{w}^L) \\
(\fullvec{r}^L\cdot\fullvec{v}^L)(\fullvec{s}^L\cdot\fullvec{u}^L)(\fullvec{t}^L\cdot\fullvec{w}^L) \\
(\fullvec{r}^L\cdot\fullvec{v}^L)(\fullvec{s}^L\cdot\fullvec{w}^L)(\fullvec{t}^L\cdot\fullvec{u}^L) \\
(\fullvec{r}^L\cdot\fullvec{w}^L)(\fullvec{s}^L\cdot\fullvec{t}^L)(\fullvec{u}^L\cdot\fullvec{v}^L) \\
(\fullvec{r}^L\cdot\fullvec{w}^L)(\fullvec{s}^L\cdot\fullvec{u}^L)(\fullvec{t}^L\cdot\fullvec{v}^L) \\
(\fullvec{r}^L\cdot\fullvec{w}^L)(\fullvec{s}^L\cdot\fullvec{v}^L)(\fullvec{t}^L\cdot\fullvec{u}^L)
\end{bmatrix}
\label{eq:rank6vec}
\end{align}
\end{subequations}
Performing the necessary operations, we obtain 
\begin{align}
b_{0,0,0,0} =& \dfrac{1}{6}\left( \sum_{\mu^M} \int d\Theta_k^M \left| \fullvec{D}_{\fullvec{k}^M,\mu^M}^M \right|^2 \right) \left| \fullvec{E}^L \right|^2 
%
%
\end{align}
\begin{align}
b_{1,0,0,0} =& \dfrac{1}{4\sqrt{3}} \left\{ \sum_{\mu^M} \int d\Theta_k^M \left[ \unitvec{k}^M \cdot \left( i \fullvec{D}_{\fullvec{k}^M,\mu^M}^{M*} \times \fullvec{D}_{\fullvec{k}^M,\mu^M}^M  \right) \right] \right\} \left[ \unitvec{z}^L \cdot \left( -i \fullvec{E}^{L*} \times \fullvec{E}^L \right) \right] \nonumber \\
=& \dfrac{\xi}{4\sqrt{3}} \left\{ \int d\Theta_k^M \left[ \unitvec{k}^M \cdot \text{Tr} \left(  \vec{\mathbb{B}}_{\fullvec{k}^M}^M  \right) \right] \right\} \left| \fullvec{E}^L \right|^2
\end{align}
\begin{align}
b_{2,0,0,0} =& \dfrac{1}{12\sqrt{5}} \left[ \sum_{\mu^M} \int d\Theta_k^M \left( \left| \fullvec{D}_{\fullvec{k}^M,\mu^M}^M \right|^2 - 3 \left| \unitvec{k}^M \cdot \fullvec{D}_{\fullvec{k}^M,\mu^M}^M \right|^2  \right) \right] \left| \fullvec{E}^L \right|^2 \nonumber \\
=& \dfrac{1}{12\sqrt{5}} \left\{  \sum_{\mu^M} \int d\Theta_k^M \left| \fullvec{D}_{\fullvec{k}^M,\mu^M}^M \right|^2 - 3 \int d\Theta_k^M \, \text{Tr}\left( \unitvec{k}^M \cdot \vec{\mathbb{K}}_{\fullvec{k}^M}^M \right) \right\} \left| \fullvec{E}^L \right|^2, 
\end{align}
\begin{align}
b_{0,0,1,0} =& \dfrac{1}{12\sqrt{3}} \left\{ \sum_{\mu_1^M,\mu_2^M} \int d\Theta_k^M \left[ \unitvec{\sigma}_{\mu_2^M,\mu_1^M} \cdot \left( i\fullvec{D}_{\fullvec{k}^M,\mu_1^M}^{M*} \times \fullvec{D}_{\fullvec{k}^M,\mu_2^M}^{M} \right) \right]  \right\} \left[ \unitvec{z}^L \cdot \left( -i \fullvec{E}^{L*} \times \fullvec{E}^L \right) \right] \nonumber \\
=& \dfrac{\xi}{12\sqrt{3}} \left[ \int d\Theta_k^M \text{Tr}\left( \unitvec{\sigma}^M \cdot \vec{\mathbb{B}}_{\fullvec{k}^M}^M \right) \right] \left| \fullvec{E}^L \right|^2
\end{align}
\begin{align}
b_{1,0,1,0} =& \dfrac{1}{15} \left\{ \sum_{\mu_1^M,\mu_2^M} \int d\Theta_k^M \left[ \left( \fullvec{D}_{\fullvec{k}^M,\mu_1^M}^{M*} \cdot \fullvec{D}_{\fullvec{k}^M,\mu_2^M}^{M} \right) \left(   \unitvec{\sigma}_{\mu_2^M,\mu_1^M} \cdot \unitvec{k}^M \right) \right] \right\} \left| \fullvec{E}^L \right|^2 \nonumber \\
&-\dfrac{1}{30} \left\{ \sum_{\mu_1^M,\mu_2^M} \int d\Theta_k^M \text{Re} \left[ \left( \fullvec{D}_{\fullvec{k}^M,\mu_1^M}^{M*} \cdot \unitvec{\sigma}_{\mu_2^M,\mu_1^M} \right) \left( \fullvec{D}_{\fullvec{k}^M,\mu_2^M}^{M}   \cdot \unitvec{k}^M \right) \right] \right\}  \left| \fullvec{E}^L \right|^2 \nonumber \\
=& \dfrac{1}{30} \left\{ \int d\Theta_k^M \, \left[ 2 \left(\unitvec{k}^M \cdot \fullvec{S}_{\fullvec{k}^M}^M \right) - \text{Tr} \left[ \text{Re}\left( \unitvec{\sigma}^M \cdot \vec{\mathbb{K}}_{\fullvec{k}^M}^M \right)  \right] \right] \right\} \left| \fullvec{E}^L \right|^2
\end{align}
\begin{align}
b_{1,1,1,-1} + b_{1,-1,1,1} =& -\dfrac{1}{10} \left\{ \sum_{\mu_1^M,\mu_2^M} \int d\Theta_k^M \left[ \left( \fullvec{D}_{\fullvec{k}^M,\mu_1^M}^{M*} \cdot \fullvec{D}_{\fullvec{k}^M,\mu_2^M}^{M} \right) \left(   \unitvec{\sigma}_{\mu_2^M,\mu_1^M} \cdot \unitvec{k}^M \right) \right] \right\} \left| \fullvec{E}^L \right|^2 \nonumber \\
&-\dfrac{1}{30} \left\{ \sum_{\mu_1^M,\mu_2^M} \int d\Theta_k^M \text{Re} \left[ \left( \fullvec{D}_{\fullvec{k}^M,\mu_1^M}^{M*} \cdot \unitvec{\sigma}_{\mu_2^M,\mu_1^M} \right) \left( \fullvec{D}_{\fullvec{k}^M,\mu_2^M}^{M}   \cdot \unitvec{k}^M \right) \right] \right\}  \left| \fullvec{E}^L \right|^2 \nonumber \\
=& -\dfrac{1}{30} \left\{ \int d\Theta_k^M \, \left[ 3 \left(\unitvec{k}^M \cdot \fullvec{S}_{\fullvec{k}^M}^M \right) + \text{Tr} \left[ \text{Re}\left( \unitvec{\sigma}^M \cdot \vec{\mathbb{K}}_{\fullvec{k}^M}^M \right)  \right] \right] \right\} \left| \fullvec{E}^L \right|^2
\end{align}
\begin{align}
b_{1,-1,1,1} - b_{1,1,1,-1} =& \dfrac{i}{12} \left\{ \sum_{\mu_1^M,\mu_2^M} \int d\Theta_k^M \left[ \left( \unitvec{k}^M \times \unitvec{\sigma}_{\mu_2^M,\mu_1^M} \right) \cdot \left( i\fullvec{D}_{\fullvec{k}^M,\mu_1^M}^{M*} \times \fullvec{D}_{\fullvec{k}^M,\mu_2^M}^{M} \right) \right]  \right\} \left[ \unitvec{z}^L \cdot \left( -i \fullvec{E}^{L*} \times \fullvec{E}^L \right) \right] \nonumber \\
=&  i\dfrac{\xi}{12} \left\{ \int d\Theta_k^M \left[ \unitvec{k}^M \cdot \text{Tr} \left( \unitvec{\sigma}^M \times \vec{\mathbb{B}}_{\fullvec{k}^M}^M \right) \right] \right\} \left| \fullvec{E}^L \right|^2
\end{align}
\begin{align}
b_{2,0,1,0} =& -\dfrac{1}{12\sqrt{15}} \left\{  \sum_{\mu_1^M,\mu_2^M} \int d\Theta_k^M \left\{ \left( \unitvec{k}^M \times \unitvec{\sigma}_{\mu_2^M,\mu_1^M}^M \right) \cdot \left[ \unitvec{k}^M \times \left( i \fullvec{D}_{\fullvec{k}^M,\mu_1^M}^{M*} \times \fullvec{D}_{\fullvec{k}^M,\mu_2^M}^{M} \right) \right]\right\} \right\} \left[ \unitvec{z}^L \cdot \left( -i \fullvec{E}^{L*} \times \fullvec{E}^L \right) \right] \nonumber \\
&+ \dfrac{1}{6\sqrt{15}} \left\{  \sum_{\mu_1^M,\mu_2^M} \int d\Theta_k^M \left\{ \left( \unitvec{k}^M \cdot \unitvec{\sigma}_{\mu_2^M,\mu_1^M}^M \right) \left[ \unitvec{k}^M \cdot \left( i \fullvec{D}_{\fullvec{k}^M,\mu_1^M}^{M*} \times \fullvec{D}_{\fullvec{k}^M,\mu_2^M}^{M} \right) \right]\right\} \right\} \left[ \unitvec{z}^L \cdot \left( -i \fullvec{E}^{L*} \times \fullvec{E}^L \right) \right] \nonumber \\
=& -\dfrac{\xi}{12\sqrt{15}} \left\{ \int d\Theta_k^M \, \text{Tr} \left[ \left(\unitvec{k}^M\times\unitvec{\sigma}^M\right)\cdot\left(\unitvec{k}^M\times\vec{\mathbb{B}}_{\fullvec{k}^M}^M\right) - 2  \left(\unitvec{k}^M\cdot\unitvec{\sigma}^M\right)\left(\unitvec{k}^M\cdot\vec{\mathbb{B}}_{\fullvec{k}^M}^M\right) \right] \right\}  \left| \fullvec{E}^L \right|^2 
\end{align}
\begin{align}
b_{2,-1,1,1} - b_{2,1,1,-1} =& -\dfrac{i}{6\sqrt{5}} \left\{  \sum_{\mu_1^M,\mu_2^M} \int d\Theta_k^M \text{Re} \left\{ \left[ \left( \unitvec{k}^M \times \unitvec{\sigma}_{\mu_2^M,\mu_1^M}^M \right) \cdot \fullvec{D}_{\fullvec{k}^M,\mu_1^M}^{M*}  \right]\left( \unitvec{k}^M \cdot \fullvec{D}_{\fullvec{k}^M,\mu_2^M}^{M}  \right) \right\} \right\}  \left| \fullvec{E}^L \right|^2 \nonumber \\
=& -\dfrac{i}{6\sqrt{5}} \left\{ \int d\Theta_k^M \, \text{Re} \left[ \unitvec{k}^M \cdot \text{Tr} \left( \unitvec{\sigma}^M \times  \vec{\mathbb{K}}_{\fullvec{k}^M}^M \right) \right] \right\}  \left| \fullvec{E}^L \right|^2
\end{align}
\begin{align}
b_{3,0,1,0} =&  \dfrac{1}{20\sqrt{21}} \left\{ \sum_{\mu_1^M,\mu_2^M} \int d\Theta_k^M \left[ \left( \fullvec{D}_{\fullvec{k}^M,\mu_1^M}^{M*} \cdot \fullvec{D}_{\fullvec{k}^M,\mu_2^M}^{M} \right) \left(   \unitvec{\sigma}_{\mu_2^M,\mu_1^M} \cdot \unitvec{k}^M \right) \right] \right\} \left| \fullvec{E}^L \right|^2 \nonumber \\
&+\dfrac{1}{10\sqrt{21}} \left\{ \sum_{\mu_1^M,\mu_2^M} \int d\Theta_k^M \text{Re} \left[ \left( \fullvec{D}_{\fullvec{k}^M,\mu_1^M}^{M*} \cdot \unitvec{\sigma}_{\mu_2^M,\mu_1^M} \right) \left( \fullvec{D}_{\fullvec{k}^M,\mu_2^M}^{M}   \cdot \unitvec{k}^M \right) \right] \right\}  \left| \fullvec{E}^L \right|^2 \nonumber \\
& - \dfrac{1}{4\sqrt{21}} \left\{ \sum_{\mu_1^M,\mu_2^M} \int d\Theta_k^M \left[ \left( \unitvec{k}^M \cdot \fullvec{D}_{\fullvec{k}^M,\mu_1^M}^{M*}  \right)\left( \unitvec{k}^M \cdot \fullvec{D}_{\fullvec{k}^M,\mu_2^M}^{M*}  \right) \left(   \unitvec{\sigma}_{\mu_2^M,\mu_1^M} \cdot \unitvec{k}^M \right)  \right] \right\}  \left| \fullvec{E}^L \right|^2 \nonumber \\
=& \dfrac{1}{20\sqrt{21}} \left\{ \int d\Theta_k^M \,  \left[ \left(\unitvec{k}^M \cdot \fullvec{S}_{\fullvec{k}^M}^M \right) + \text{Tr} \left[ 2 \text{Re}\left( \unitvec{\sigma}^M \cdot \vec{\mathbb{K}}_{\fullvec{k}^M}^M \right) -5\left(\unitvec{\sigma}^M\cdot\unitvec{k}^M\right)\left( \unitvec{k}^M \cdot \vec{\mathbb{K}}_{\fullvec{k}^M}^M \right) \right]  \right] \right\}  \left| \fullvec{E}^L \right|^2 
\end{align}
wherein, we introduced the following quantities
\begin{align}
\left( \vec{\mathbb{B}}_{\fullvec{k}^M}^M \right)_{\mu_1^M,\mu_2^M} = i \fullvec{D}_{\fullvec{k}^M,\mu_1^M}^{M*} \times \fullvec{D}_{\fullvec{k}^M,\mu_2^M}^{M}
\end{align}
\begin{align}
\fullvec{S}_{\fullvec{k}^M}^M = \sum_{\mu_1^M,\mu_2^M} \left( \fullvec{D}_{\fullvec{k}^M,\mu_1^M}^{M*} \cdot \fullvec{D}_{\fullvec{k}^M,\mu_2^M}^{M} \right) \unitvec{\sigma}_{\mu_2^M,\mu_1^M}
\end{align}
\begin{align}
\left( \vec{\mathbb{K}}_{\fullvec{k}^M}^M \right)_{\mu_1^M,\mu_2^M} =  \fullvec{D}_{\fullvec{k}^M,\mu_1^M}^{M*} \left( \unitvec{k}^M \cdot \fullvec{D}_{\fullvec{k}^M,\mu_2^M}^{M}  \right)
\end{align}
and $\xi=\pm1$ is a dichroic parameter characterizing the rotation of the light polarization vector. We did not list the explicit form for the coefficients $b_{2,\pm1,1,\mp1},b_{3,\pm1,1,\mp1}$ since these also satisfy Eqs. \eqref{eq:rel1} and \eqref{eq:rel2}. The photoionization parameters Eq. \eqref{eq:dynamic} are then simply obtained by substituting the coefficients $b_{\ell,m_\ell,\ell_s,m_s}$ into Eq. \eqref{eq:parameters}. 

\section{Recovering Cherepkov's expression}
\label{app:recover}

Recall that Cherepkov's expansion coefficients for the photoionization yield are given as 
\begin{align}
A_{L,M_{L},S,M_{S}}= & \dfrac{4\pi \sqrt{2\pi}}{3}  |E_\omega^L|^{2}  \sum(-1)^{m_{2}'+\xi'-\xi+\mu_{2}'-1/2} (2J+1) \sqrt{\frac{(2\ell_1+1)(2\ell_2+1)(2L+1)}{4\pi}} \nonumber \\
&\times 
\begin{pmatrix}
	\ell_{2} & \ell_{1} & L\\
	0 & 0 & 0
\end{pmatrix}
\begin{pmatrix}
	\ell_{2} & \ell_{1} & L\\
	m_{2}' & -m_{1}' & M_{L}'
\end{pmatrix} 
\begin{pmatrix}
	1 & 1 & J\\
	\xi'' & -\xi' & -M_{J}'
\end{pmatrix}
\begin{pmatrix}
	1 & 1 & J\\
	\xi & -\xi & 0
\end{pmatrix} \nonumber \\
&\times 
\begin{pmatrix}
	\frac{1}{2} & \frac{1}{2} & S\\
	\mu_{2}' & -\mu_{1}' & M_{S}'
\end{pmatrix}
\begin{pmatrix}
	L & S & J\\
	M_{L}' & M_{S}' & M_{J}'
\end{pmatrix}
\begin{pmatrix}
	L & S & J\\
	-M_{L} & -M_{S} & 0
\end{pmatrix}
(D_{\xi'}^{\ell_1,m_1',\mu_1'})^* D_{\xi''}^{\ell_2,m_2',\mu_2'}.
\label{eq:app_cherepkovcoeff}
\end{align}
It will be non-trivial to prove that $A_{L,M_L,S,M_S}=b_{L,M_L,S,M_S}$ for all non-zero values of $A_{L,M_L,S,M_S}$. Here, we shall only explicitly show that we can use $A_{L,M_L,S,M_S}$, i.e., 
\begin{align}
D=&\frac{\sqrt{3}A_{1,0,0,0}}{A_{0,0,0,0}} \label{eq:app_cherepkov-D}
\end{align}
\begin{align}
A =& \frac{\sqrt{3}A_{0,0,1,0}}{A_{0,0,0,0}} \label{eq:app_cherepkov-A}
\end{align}
\begin{align}
C =& -i \frac{3 (A_{1,-1,1,1}-A_{1,1,1,-1})}{2 A_{0,0,0,0}} \label{eq:app_cherepkov-C}
\end{align}
and recover the expressions for the parameters $\{D,A,C\}$ in Eq. \eqref{eq:dynamic}. This direct equivalence together with $b_{2,\pm1,1,\mp1},b_{3,\pm1,1,\mp1}$ satisfying Eqs. \eqref{eq:rel1} and \eqref{eq:rel2} is then enough to provide confidence on the equivalence of the two approaches. In the succeeding expressions, we shall use the spherical components of a vector (see for example Ref. \cite{varshalovich1988quantum}),
\begin{align}
V_0 = V_z \quad , \quad V_\pm = \mp \dfrac{1}{\sqrt{2}}(V_x\pm i V_y)
\end{align}
such that the dot and scalar products can be written as 
\begin{align}
\fullvec{A}\cdot\fullvec{B} =& \sum_q (-1)^q A_q B_{-q}
\label{eq-app:dot}
\end{align}
\begin{align}
(\fullvec{A}\times\fullvec{B})_r =& -i \sqrt{6}(-1)^r 
\begin{pmatrix}
    1 & 1 & 1 \\
    p & q & -r
\end{pmatrix}
A_p B_q.
\label{eq-app:cross}
\end{align}

We first consider the normalizing factor $A_{0,0,0,0}$, which upon evaluating the $3j-$symbols will simplify to
\begin{align}
A_{0,0,0,0} = & \dfrac{4\pi\sqrt{\pi}}{9}|E_{\omega}^L|^{2}\sum(-1)^{m_{1}'}  \sqrt{\frac{(2\ell_1+1)(2\ell_2+1) }{4\pi}}
\begin{pmatrix}
\ell_{2} & \ell_{1} & 0\\
0 & 0 & 0
\end{pmatrix}
\begin{pmatrix}
\ell_{2} & \ell_{1} & 0\\
m_{2}' & -m_{1}' & 0
\end{pmatrix}
(D_{\xi'}^{\ell_{1},m_{1}',\mu_{1}'})^* D_{\xi'}^{\ell_{2},m_{2}',\mu_{1}'} \nonumber \\
=& \dfrac{2\pi}{9}|E_{\omega}^L|^{2}\sum \int d\Theta_k^M  (-1)^{\xi'} (\fullvec{D}^{\ell_{1},m_{1}',\mu_{1}'*})_{-\xi'} D_{\xi'}^{\ell_{2},m_{2}',\mu_{1}'}  Y_{\ell_1,m_1'}^*(\unitvec{k}^M) Y_{\ell_2,m_2'}(\unitvec{k}^M) \nonumber \\
=& \dfrac{2\pi}{9} |E_\omega^L|^2 \sum \int d\Theta_k^M Y_{\ell_1,m_1'}^*(\unitvec{k}^M) Y_{\ell_2,m_2'}(\unitvec{k}^M) \left[ \fullvec{D}^{\ell_{1},m_{1}',\mu_{1}'*} \cdot \fullvec{D}^{\ell_{2},m_{2}',\mu_{1}'} \right]  \nonumber \\
=& \dfrac{1}{6}|E_{\omega}^L|^{2} \sum_{\mu^M=\pm\frac{1}{2}} \int d\Theta_k^M  \left| \fullvec{D}^M_{\fullvec{k}^M,\mu^M} \right|^2
\label{eq:app_A0000_vec}
\end{align}
The second line follows from the following definitions 
\begin{align}
(V_q)^*=(-1)^q(\fullvec{V}^*)_{-q}
\end{align}
\begin{align}
\int d\Theta_k Y_{\ell_1,m_1'}(\unitvec{k})Y_{\ell_2,m_2'}(\unitvec{k})Y_{L,M_L'}(\unitvec{k}) = \sqrt{ \dfrac{(2\ell_1+1)(2\ell_2+1)(2L+1)}{4\pi} } \begin{pmatrix}
	\ell_{2} & \ell_{1} & L\\
	0 & 0 & 0
\end{pmatrix}
\begin{pmatrix}
	\ell_{2} & \ell_{1} & L\\
	m_{2}' & m_{1}' & M_{L}'
\end{pmatrix},
\end{align}
while the third line follows from Eq. \eqref{eq-app:dot}. The last line follows from the definition of the dipole matrix elements Eq. \eqref{eq:photoelectron}, i.e., 
\begin{align}
\fullvec{D}^M_{\fullvec{k}^M,\mu^M} = \langle \Psi_{\fullvec{k}^M,\mu^M}^{(-)} | \fullvec{d}^M | \psi_i \rangle =  \sum_{\ell,m} \bigg \langle \varphi_{\ell,m,\mu^M}^{(-)} Y_{\ell,m}^*(\unitvec{k}^M) \bigg| \sqrt{\dfrac{4\pi}{3}} r Y_{1,q} \unitvec{e}_q \bigg| \psi_i \bigg\rangle = \sqrt{\dfrac{4\pi}{3}}\sum_{\ell,m} \fullvec{D}^{\ell,m,\mu} Y_{\ell,m}(\unitvec{k}^M)
\end{align}

Repeating the same steps for $A_{1,0,0,0}$, we get
\begin{align}
A_{1,0,0,0} =& \dfrac{4\pi}{3}\sqrt{\dfrac{\pi}{6}} \xi  |E_{\omega}^L|^{2}\sum (-1)^{m_{1}'+\xi'}  \sqrt{\frac{3(2\ell_1+1)(2\ell_2+1) }{4\pi}}
\begin{pmatrix}
\ell_{2} & \ell_{1} & 1\\
0 & 0 & 0
\end{pmatrix}
\begin{pmatrix}
\ell_{2} & \ell_{1} & 1\\
m_{2}' & -m_{1}' & M_L'
\end{pmatrix}
\begin{pmatrix}
1 & 1 & 1 \\
-\xi' & \xi'' & M_L'
\end{pmatrix} \nonumber \\
& (D_{\xi'}^{\ell_{1},m_{1}',\mu_{1}'})^* D_{\xi'}^{\ell_{2},m_{2}',\mu_{1}'} \nonumber \\
=& \dfrac{4\pi}{3}\sqrt{\dfrac{\pi}{6}} \xi  |E_{\omega}^L|^{2} \sum \int d\Theta_k^M  
\begin{pmatrix}
1 & 1 & 1 \\
-\xi' & \xi'' & M_L'
\end{pmatrix}
(\fullvec{D}^{\ell_{1},m_{1}',\mu_{1}'*})_{-\xi'} D_{\xi'}^{\ell_{2},m_{2}',\mu_{1}'} Y_{1,M_L'}(\unitvec{k}^M)  Y_{\ell_1,m_1'}^*(\unitvec{k}^M) Y_{\ell_2,m_2'}(\unitvec{k}^M) \nonumber \\
=& \dfrac{\sqrt{\pi}}{6} \xi |E_{\omega}^L|^{2} \sum \int d\Theta_k^M  (-1)^{M_L'} Y_{1,M_L'}(\unitvec{k}^M)  \left( i \fullvec{D}_{\fullvec{k}^M,\mu^M}^{M*} \times \fullvec{D}_{\fullvec{k}^M,\mu^M}^M\right)_{-M_L'} \nonumber \\
=& \dfrac{1}{4\sqrt{3}} \xi |E_{\omega}^L|^{2} \left\{ \sum_{\mu^M} \int d\Theta_k^M \left[ \unitvec{k}^M \cdot \left( i \fullvec{D}_{\fullvec{k}^M,\mu^M}^{M*} \times \fullvec{D}_{\fullvec{k}^M,\mu^M}^M  \right) \right] \right\} \nonumber \\
=& \dfrac{1}{4\sqrt{3}} \xi |E_{\omega}^L|^{2} \int d\Theta_k^M \left[ \unitvec{k}^M \cdot \text{Tr}(\vec{\mathbb{B}}_{\fullvec{k}^M}^M) \right]
\label{eq:app_A1000_vec}
\end{align}
Substituting Eqs. \eqref{eq:app_A0000_vec} and \eqref{eq:app_A1000_vec} into Eq. \eqref{eq:app_cherepkov-D} we thus obtain 
\begin{align}
D=&\frac{\sqrt{3}A_{1,0,0,0}}{A_{0,0,0,0}}  = \dfrac{3}{2 S_0} \int d\Theta_k^M \left[ \unitvec{k}^M \cdot \text{Tr}(\vec{\mathbb{B}}_{\fullvec{k}^M}^M) \right]
\end{align}
\begin{align}
S_0 =&  \sum_{\mu^M=\pm\frac{1}{2}} \int d\Theta_k^M  \left| \fullvec{D}^M_{\fullvec{k}^M,\mu^M} \right|^2
\end{align}
which is exactly equal to Eq. \eqref{eq:para-D-gen}.

Next, we have $A_{0,0,1,0}$ which simplifies to 
\begin{align}
A_{0,0,1,0} =& \dfrac{4\pi}{3} \sqrt{\dfrac{\pi}{3}} \xi |E_{\omega}^L|^{2} \sum (-1)^{M_S'+m_1'+\xi'+\mu_2'+-1/2} \sqrt{\dfrac{(2\ell_1+1)(2\ell_2+1)}{4\pi}} 
\begin{pmatrix}
\ell_1 & \ell_2 & 0 \\
0 & 0 & 0
\end{pmatrix}
\begin{pmatrix}
\ell_2 & \ell_1 & 0 \\
m_2' & -m_1 ' & 0
\end{pmatrix} \nonumber \\
& 
\begin{pmatrix}
1 & 1 & 1 \\
-\xi' & \xi'' & M_S'
\end{pmatrix}
\begin{pmatrix}
\frac{1}{2} & \frac{1}{2} & 1 \\
\mu_2' & - \mu_1' & M_S'
\end{pmatrix}
(D_{\xi'}^{\ell_{1},m_{1}',\mu_{1}'})^* D_{\xi'}^{\ell_{2},m_{2}',\mu_{1}'}  \nonumber \\
=& \dfrac{2\pi}{3\sqrt{3}} \xi |E_{\omega}^L|^{2} \sum \int d\Theta_k^M (-1)^{\mu_2'-1/2} 
\begin{pmatrix}
\frac{1}{2} & \frac{1}{2} & 1 \\
\mu_2' & - \mu_1' & M_S'
\end{pmatrix} \nonumber \\
& (-1)^{M_S'} 
\begin{pmatrix}
1 & 1 & 1 \\
-\xi' & \xi'' & M_S'
\end{pmatrix}
(\fullvec{D}^{\ell_{1},m_{1}',\mu_{1}'*})_{-\xi'} D_{\xi'}^{\ell_{2},m_{2}',\mu_{1}'}Y_{\ell_1,m_1'}^*(\unitvec{k}^M) Y_{\ell_2,m_2'}(\unitvec{k}^M) \nonumber \\
=& \dfrac{1}{6\sqrt{2}} \xi |E_{\omega}^L|^{2} \sum \int d\Theta_k^M (-1)^{\mu_2'-1/2} 
\begin{pmatrix}
\frac{1}{2} & \frac{1}{2} & 1 \\
\mu_2' & - \mu_1' & M_S'
\end{pmatrix}
\left( i \fullvec{D}_{\fullvec{k}^M,\mu_1'}^{M*} \times  \fullvec{D}_{\fullvec{k}^M,\mu_2'}^{M} \right)_{-M_S'} \nonumber \\
=& \dfrac{1}{12\sqrt{3}} \xi |E_{\omega}^L|^{2} \sum \int d\Theta_k^M \left[ \unitvec{\sigma}_{\mu_2^M,\mu_1^M} \cdot \left( i\fullvec{D}_{\fullvec{k}^M,\mu_1^M}^{M*} \times \fullvec{D}_{\fullvec{k}^M,\mu_2^M}^{M} \right) \right] \nonumber \\
=& \dfrac{1}{12\sqrt{3}} \xi |E_{\omega}^L|^{2}  \int d\Theta_k^M \, \text{Tr}\left( \unitvec{\sigma}^M \cdot \vec{\mathbb{B}}_{\fullvec{k}^M}^M \right)
\end{align}
which together with Eq. \eqref{eq:app_A0000_vec} is exactly equal to Eq. \eqref{eq:para-D-gen}. The last line follows from expanding the summation over $M_S'$ and rewriting the covariant spherical basis vectors in terms of $\unitvec{\sigma}_{\mu_2,\mu_1}$.   

Last, we have 
\begin{align}
A_{1,-1,1,+1} - A_{1,+1,1,-1} =& \dfrac{4\pi\sqrt{2\pi}}{3} \xi |E_{\omega}^L|^{2} \sum (-1)^{m_1'+M_L'+\xi'+\mu_2'-1/2} \sqrt{\dfrac{3(2\ell_1+1)(2\ell_2+1)}{4\pi}} \nonumber \\
&
\begin{pmatrix}
\ell_1 & \ell_2 & 1 \\
0 & 0 & 0
\end{pmatrix}
\begin{pmatrix}
\ell_2 & \ell_1 & 1 \\
m_2' & -m_1 ' & M_L'
\end{pmatrix}
\begin{pmatrix}
1 & 1 & 1 \\
-\xi' & \xi'' & -M_J'
\end{pmatrix}
\begin{pmatrix}
\frac{1}{2} & \frac{1}{2} & 1 \\
\mu_2' & - \mu_1' & M_S'
\end{pmatrix} \nonumber \\
&
\begin{pmatrix}
1 & 1 & 1 \\
M_L' & M_S' & M_J'
\end{pmatrix}
(D_{\xi'}^{\ell_{1},m_{1}',\mu_{1}'})^* D_{\xi'}^{\ell_{2},m_{2}',\mu_{1}'} \nonumber \\
=& \dfrac{4\pi\sqrt{2\pi}}{3} \xi |E_{\omega}^L|^{2} \sum \int d\Theta_k^M (-1)^{\mu_2'-1/2}
\begin{pmatrix}
\frac{1}{2} & \frac{1}{2} & 1 \\
\mu_2' & - \mu_1' & M_S'
\end{pmatrix}
(-1)^{M_L'}
\begin{pmatrix}
1 & 1 & 1 \\
M_L' & M_S' & M_J'
\end{pmatrix}
Y_{1,M_L'}(\unitvec{k}^M) \nonumber \\
& 
\begin{pmatrix}
1 & 1 & 1 \\
-\xi' & \xi'' & -M_J'
\end{pmatrix}
(\fullvec{D}^{\ell_{1},m_{1}',\mu_{1}'*})_{-\xi'} D_{\xi'}^{\ell_{2},m_{2}',\mu_{1}'} Y^*_{\ell_1,m_1}(\unitvec{k}^M) Y_{\ell_2,m_2} (\unitvec{k}^M) \nonumber \\
=& -\dfrac{i}{2\sqrt{6}} \xi |E_{\omega}^L|^{2} \sum \int d\Theta_k^M (-1)^{\mu_2'-1/2}
\begin{pmatrix}
\frac{1}{2} & \frac{1}{2} & 1 \\
\mu_2' & - \mu_1' & M_S'
\end{pmatrix} \left[ \unitvec{k}^M \times \left( i\fullvec{D}_{\fullvec{k}^M,\mu_1'}^{M*} \times \fullvec{D}_{\fullvec{k}^M,\mu_2'}^{M} \right)  \right] \nonumber \\
=& \dfrac{i}{12}  \xi |E_{\omega}^L|^{2} \int d\Theta_k^M \left[ \unitvec{k}^M \cdot \text{Tr} \left( \unitvec{\sigma}^M \times \vec{\mathbb{B}}_{\fullvec{k}^M}^M \right) \right]
\end{align}
which is exactly equal to Eq. \eqref{eq:para-C-gen} after normalizing with Eq. \eqref{eq:app_A0000_vec}.


\end{widetext}

\bibliographystyle{apsrev4-2}
\bibliography{spin-current}

\end{document}